\documentclass[]{aastex631}

\newcommand\kms{km~s$^{-1}$}
\defcitealias{yuan2017high}{Y17}
\usepackage{multirow}
\begin{document}

\title{High-Mass Starless Clumps: Dynamical State and Correlation Between Physical Parameters}

\author[0000-0001-7393-8583]{Bo Huang}
\affiliation{Guangxi Key Laboratory for Relativistic Astrophysics, School of Physical Science and Technology, Nanning, 530004, China}
\affiliation{Kavli Institute for Astronomy and Astrophysics, Peking University, Haidian District, Beijing, 100871, China}
\affiliation{Institut de Ciències de l’Espai (ICE), CSIC, Can Magrans s/n, E-08193 Cerdanyola del Vallès, Catalonia, Spain}
\affiliation{Institut d’Estudis Espacials de Catalunya (IEEC), E-08034 Barcelona, Catalonia, Spain}

\author{Ke Wang}
\altaffiliation{Corresponding author}
\email{kwang.astro@pku.edu.cn}
\affiliation{Kavli Institute for Astronomy and Astrophysics, Peking University, Haidian District, Beijing, 100871, China}

\author{Josep M. Girart}
\affiliation{Institut de Ciències de l’Espai (ICE), CSIC, Can Magrans s/n, E-08193 Cerdanyola del Vallès, Catalonia, Spain}
\affiliation{Institut d’Estudis Espacials de Catalunya (IEEC), E-08034 Barcelona, Catalonia, Spain}

\author{Wenyu Jiao}
\affiliation{Kavli Institute for Astronomy and Astrophysics, Peking University, Haidian District, Beijing, 100871, China}
\affiliation{Department of Astronomy, School of Physics, Peking University, Beijing, 100871, China}

\author{Qianru He}
\affiliation{Purple Mountain Observatory, Chinese Academy of Sciences, Nanjing, 210023, China}
\affiliation{School of Astronomy and Space Sciences, University of Science and Technology of China, Hefei, 230026, China}

\author{Enwei Liang}
\affiliation{Guangxi Key Laboratory for Relativistic Astrophysics, School of Physical Science and Technology, Nanning, 530004, China}

\begin{abstract}
In order to study the initial conditions of massive star formation, we have previously built a sample of 463 high-mass starless clumps (HMSCs) across the inner Galactic plane covered by multiple continuum surveys.
Here, we use $^{13}$ CO(2-1) line data from the SEDIGISM survey, which covers 78$^{\circ}$ in longitude ($-60^{\circ}<l<18^{\circ}$, $\vert b\vert<0.5^{\circ}$) with 30$^{\prime \prime}$ resolution, to investigate the global dynamical state of these parsec-scale HMSCs (207 sources with good quality data, mass $10^{2}\sim 10^{5}\ \rm M_{\odot}$, size $0.1\sim3.6$ pc).
We find that most HMSCs are highly turbulent with a median Mach number $\mathcal{M_{S}}\sim$ 8.2, and 44\%$\sim$55\% of them are gravitationally bound (with virial parameter $\alpha_{\rm vir} \lesssim 2$) if no magnetic fields were present.
A median magnetic field strength of 0.33$\sim$0.37 mG would be needed to support these bound clumps against collapse, in agreement with previous observations of magnetic fields in massive star formation regions.
Luminosity-to-mass ratio, an important tracer for evolutionary stage, is strongly correlated with dust temperature.
Magnetic field strength is also correlated with density. The Larson linewidth-size scaling does not hold in HMSCs.
This study advances our understanding of global properties of HMSCs, and our high-resolution ALMA observations are on the way to study the resolved properties.
\end{abstract}


\section{Introduction} \label{sec:intro}
The formation of massive stars ($\geq 8\ \rm M_{\odot}$) is one of the hot topics in modern astrophysics, due to the important role that these objects play in the evolution of the Galactic ecosystem.
However, the formation of massive stars is still poorly understood because of their large distance, high extinction, and short timescales of critical evolutionary phases \citep[e.g.,][]{mckee2007theory, motte2018high}.
Two main classes of theory have been proposed to explain the formation of massive stars: \textit{core accretion} and \textit{competitive accretion}.
The \textit{core accretion} model suggests that cores of dense gas condense from clump fragmentation undergo gravitational collapse to form an individual star or a multiple system \citep{mckee2003formation}.
While in \textit{competitive accretion} model, a massive protostar gains material from the clump without being in a massive starless bound core \citep{bonnell2001competitive}.
\textit{``Global hierarchical collapse''} is an alternative scenario proposed by \cite{vazquez2019global}. In this scenario, high-mass protostars form within low-mass prestellar cores which grow from low- to high-mass where global collapse drives inflowing gas streams toward the protostars. 
Therefore, the high-mass stars precursors could be low-mass cores within massive infalling clumps.
\cite{padoan2020origin} proposed an alternative model based on self-consistent and extensive simulations: the \textit{inertial-flow} model.
In this case, massive stars in general do not form from the collapse of massive cores nor from competitive accretion, but they are assembled by large-scale, converging, inertial flows naturally occur in supersonic turbulence.
Hence, to unveil the formation mechanisms and better understand the formation processes of massive stars, it is important to investigate the initial conditions of the earliest stages of high-mass star-forming regions.

Observationally, massive stars are known to form predominantly in clusters, typically in parsec-scale dense clumps embedded in giant molecular clouds. 
The initial conditions are defined in high-mass starless clumps (HMSCs), which are deemed to form high-mass stars, but has not yet started to form.
Gravitationally unstable clumps collapse and may fragment into cores (typical size $\sim 0.1$\,pc) that subsequently contract to form single star or bound systems of protostars \citep[e.g.,][]{beuther2006formation, wang2011hierarchical,wang2012protostellar,wang2014hierarchical,zhang2015fragmentation,pillai2019massive}.

Best efforts have been made by several groups to identify reliable HMSCs \citep{tackenberg2012search,traficante2015initial,svoboda2016bolocam,yuan2017high}.
The basic idea is to start with a catalog of clumps identified from blind surveys, and remove sources which already show signatures of star formation.
It is worth to note that, usually the term ``clump'' refers to pc scale round structures, but when it comes to Galactic continuum surveys (e.g., ATLASGAL, HiGAL), ``clump'' is often used to refer sources decomposed from such continuum images. That is only for practical convenience since distance (and so physical size) is not yet available.
Typical physical sizes of ATLASGAL and HiGAL ``clumps'' are indeed of the order of $\sim 1$\,pc, but with a large range \citep{csengeri2014atlasgal,elia2017hi}.

A brief overview of past efforts in identifying HMSCs is given here.
\cite{tackenberg2012search} explores the physical properties of ATLASGAL sources within the region $10^{\circ}<l<20^{\circ}$ and sorts out 210 starless clumps based on the absence of young stellar objects (YSOs) in sources from the catalogs of Galactic Legacy Infrared Mid-Plane Survey \cite[GLIMPSE, ][]{churchwell2009spitzer} and the Galactic Plane Survey using the Multiband Infrared Photometer for Spitzer aboard the Spitzer Space Telescope \citep[MIPSGAL, ][]{carey2009mipsgal}.
\cite{traficante2015initial} screens out 667 starless clumps in the Herschel infrared Galactic Plane Survey \citep[Hi-GAL, ][]{molinari2010hi} within the range of $15^{\circ}<l<55^{\circ}$ which are not associated with 70 $\mu$m bright sources.
\cite{svoboda2016bolocam} identifies 2223 starless clump candidates within $10^{\circ}<l<65^{\circ}$ based on Bolocam Galactic Plane Survey \cite[BGPS, ][]{aguirre2010bolocam} where a series of observational signatures of star formation are excluded.

In \citet{yuan2017high} (hearafter \citetalias{yuan2017high}), we present 463 HMSCs in the inner Galactic plane ($\vert l\vert<60^{\circ}$, $\vert b\vert<1^{\circ}$) by cross-matching the APEX Telescope Large Area Survey of the GALaxy \citep[ATLASGAL, ][]{schuller2009atlasgal} catalog with star formation indicators, and selecting massive clumps without known sign of star formation.
The series of selection criteria of HMSCs from \citetalias{yuan2017high} are as follows: firstly, in order to make sure that the column density is sufficient to form a high mass star, only the clumps with an 870 $\mu$m peak intensity higher than 0.5 Jy\,beam$^{-1}$ have been considered.
Then, the authors have used the SIMBAD database to search for a wide range of associated star-formation-related phenomena.
Among these clumps, those associated with YSOs in catalogs of GLIMPSE, MIPSGAL 24 $\mu$m data, and Hi-GAL 70 $\mu$m data are removed.
Finally, the remaining candidates are visually inspected with Spitzer \citep{werner2004spitzer} images to remove sources with extended infrared emission.
Recently, \cite{yang2022sedigism} searched for molecular outflows from SEDIGISM \citep{schuller2017sedigism, schuller2021sedigism}, and 55 sources in \citetalias{yuan2017high}'s sample are associated with outflows.
Thus, these sources with outflow signature should also be excluded.

The previous efforts provide a valuable catalog to systematically study HMSCs and constrain their physical properties to understand of how cluster formation is initiated and of the ensuing protocluster evolution.
Strictly speaking, those HMSCs, which are identified from single-dish shallow surveys, should only be regarded as candidates, because their potential to form massive stars needs to be further investigated by their dynamical state, for which linewidth is key information \citep[e.g.,][]{wang2009relation, zhang2015fragmentation} but is largely unavailable from the literature.
Moreover, when observed at high-resolution and high-sensitivity, some HMSCs may well contain active star-forming activities, and thus ruling out their pristine starless nature \citep[e.g.,][]{jiao2023fragmentation}.

In this work, we investigate the global dynamical properties of a Galaxy-wide catalog of well characterized HMSCs selected by \citetalias{yuan2017high}, excluding 55 sources with outflows identified by \cite{yang2022sedigism}.
The range of distance, size, and mass of the sample are 1.3$\sim$18.3 kpc, 0.1$\sim$3.6 pc, and $2.0\times10^{1}\sim 6.1\times10^{5}\ \rm M_{\odot}$, respectively \citepalias{yuan2017high}. 
Using $^{13}$CO (2-1) linewidth from the SEDIGISM survey, we aim to find which, and which fraction, of the HMSCs are going to collapse to form stars, and to better understand the role of turbulence and magnetic fields in the earliest stages of high-mass star formation (HMSF).
This systematic characterization of global properties provides robust genuine starless clumps for follow-up high-resolution observations to resolve the initial conditions for massive star formation, which we are currently carrying out with ALMA and other interferometers.
This paper is organized as follows: Section \ref{Data and method} introduces the data and method used in this work; we then present the results of dynamic properties in Section \ref{DP}; Section \ref{Disc} discusses the correlation between the physical parameters of HMSCs; finally, we give conclusions in Section \ref{conclusions}.

\section{Data and Method} \label{Data and method}

The SEDIGISM survey \citep{schuller2017sedigism,schuller2021sedigism, duarte2021sedigism} covers 78$^{\circ}$ longitude of the inner Galaxy ($-60^{\circ}<l<18^{\circ}$, $\vert b\vert<0.5^{\circ}$) with a 30$^{\prime \prime}$ resolution in $^{13}$CO (2-1) and C$^{18}$O (2-1) rotational transitions.
These two lines are usually optically thin in the Galactic interstellar medium and well suited to trace the dense molecular gas.
We use the $^{13}$CO (2-1) emission line data at the position of all the identified HMSC sources reported by \citetalias{yuan2017high} (we have also checked C$^{18}$O (2-1), but the signal is too weak to identify the line for most sources).
We then extracted spectra of the filtered sources from data cubes of SEDIGISM survey.
The extracted spectra were fitted with Gaussian profiles by using the Levenberg–Marquardt algorithm provided in the Python package \textit{lmfit} \citep{newville2016lmfit}.

We set [$V_{\rm LSR}-40$, $V_{\rm LSR}+40$] \kms as the fitting velocity range (here $V_{\rm LSR}$ is the targeted centroid velocity reported from \citetalias{yuan2017high}).
In some cases, certain spectral components may not affect the targeted component or may overlap with it, causing difficulty in fitting.
We thus adjust the fitting range by using a narrower range to exclude irrelevant components or a wider range to include additional influential components to the sources.
To ensure that the quality of filtered data was good and the reliability of the fitting results, we used the following criteria:
firstly, we discard the sources where the centroid velocity $V_{\rm LSR}$ of the component(s) have an offset larger than 5 \kms with respect to the $V_{\rm LSR}$ reported by \citetalias{yuan2017high} (the $V_{\rm LSR}$ of \citetalias{yuan2017high}'s HMSCs sample is measured from CO, NH$_{3}$, CS, etc).
For the rest sources:
(1) if only one component was detected, then the mean value of the component was assigned to the source; 
(2) if multi-components were detected, then the closest value to the $V_{\rm LSR}$ value given by \citetalias{yuan2017high} was selected and assigned to the source.
For the case of double peaks near the targeted $V_{\rm LSR}$ (within 1 \kms), we firstly check the  C$^{18}$O lines from SEDIGISM, as well as NH$_{3}$, HCO$^{+}$ and N$_{2}$H$^{+}$ from the literature \citep{wienen2012ammonia, shirley2013bolocam} to determine if the source exhibits a double-component structure.
If the source shows a double-component structure but is not covered in the aforementioned molecular lines, we initially fit it with a single component.
If the derived virial parameter of this component is $\sim$2, this candidate is likely in a collapsing state, indicating that the double peak may be due to self-absorption; otherwise, we re-fit the source with two components.
Additionally, it is crucial to ensure that the residual is flat during the fitting process.

Large offset (off-source positions or strong baseline ripples) and bad-quality (poor reliability fitting results of $\Delta V_{\rm LSR}>5$ \kms) sources were discarded, as well as the sources with large uncertainty ($\rm e_{V_{\rm LSR}}>$1 \kms).
In view of the above steps, 201 sources have been discarded, 207 sources are valid.
Among the valid sources, 41 clumps have one velocity component; 30 clumps have two components; 51 clumps have three components; 32 clumps have four components; and 53 clumps have more than 4 components.
The following study is based on these 207 sources.


\section{Analysis: Dynamical Properties} \label{DP}

\begin{figure*}
	\centering
	\includegraphics[clip=true,trim=2cm 0cm 4cm 1cm,width=0.49\textwidth]{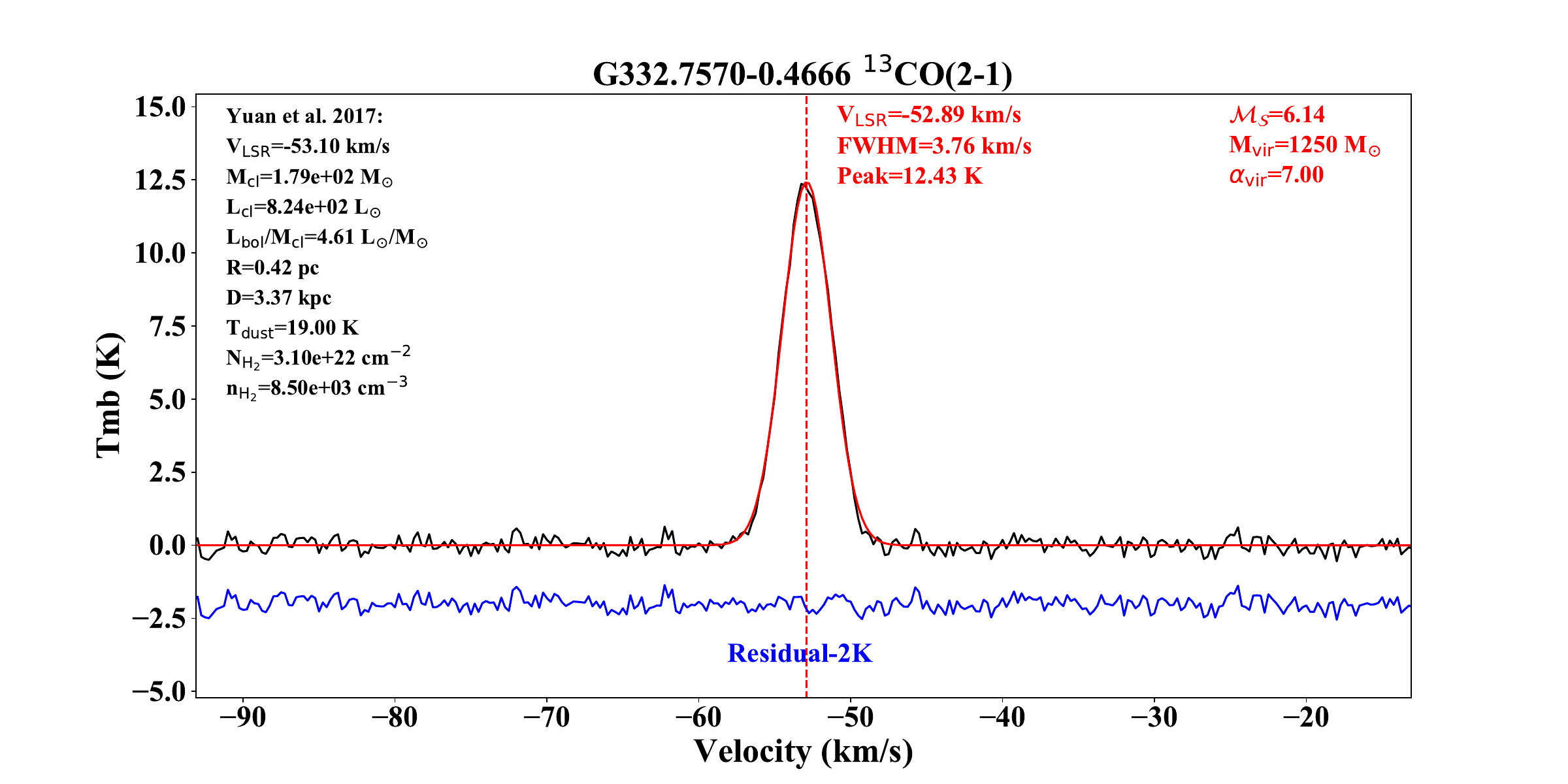}
	\includegraphics[clip=true,trim=2cm 0cm 4cm 1cm,width=0.49\textwidth]{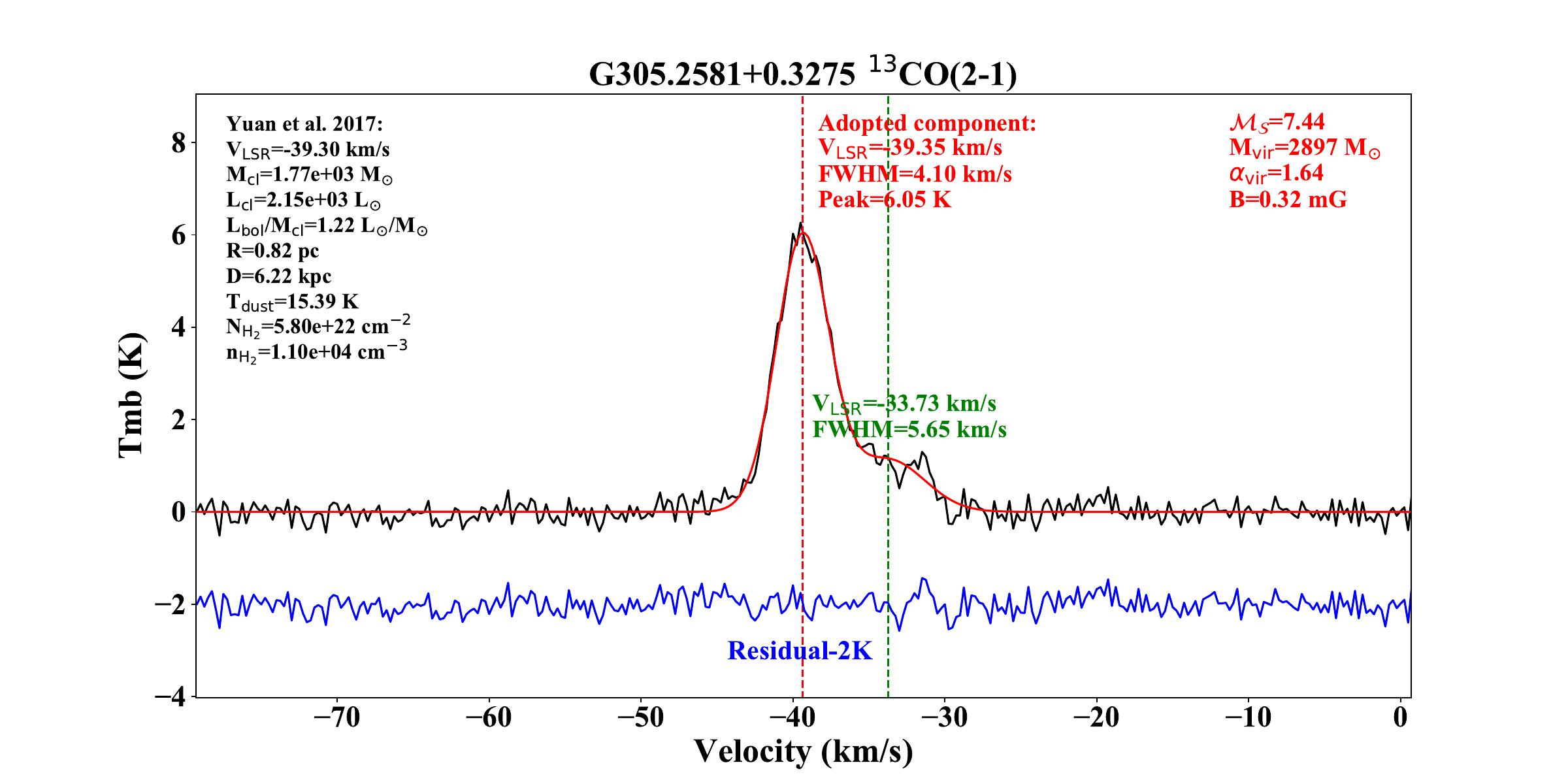}
	\includegraphics[clip=true,trim=2cm 0cm 4cm 1cm,width=0.49\textwidth]{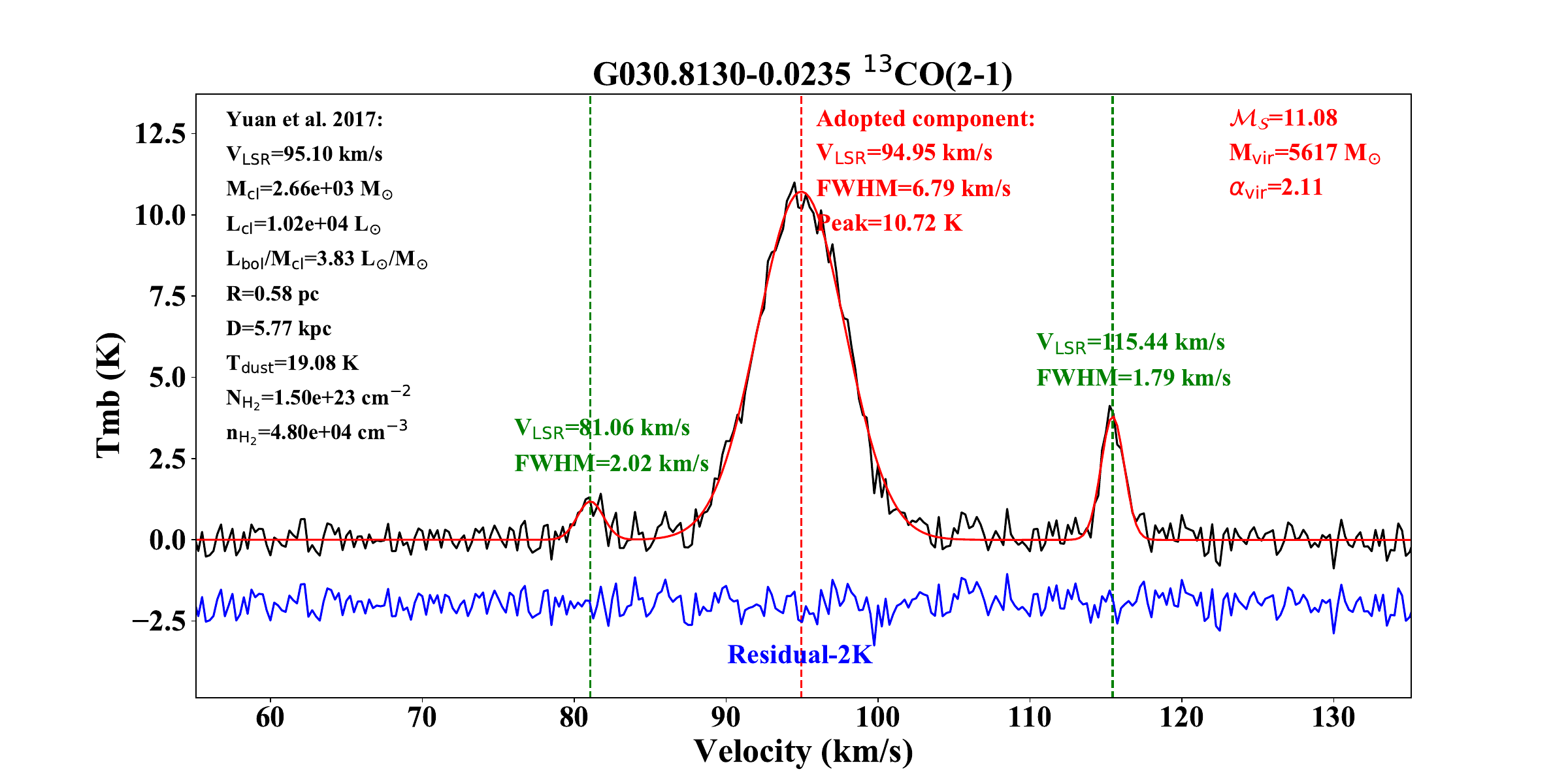}
	\includegraphics[clip=true,trim=2cm 0cm 4cm 1cm,width=0.49\textwidth]{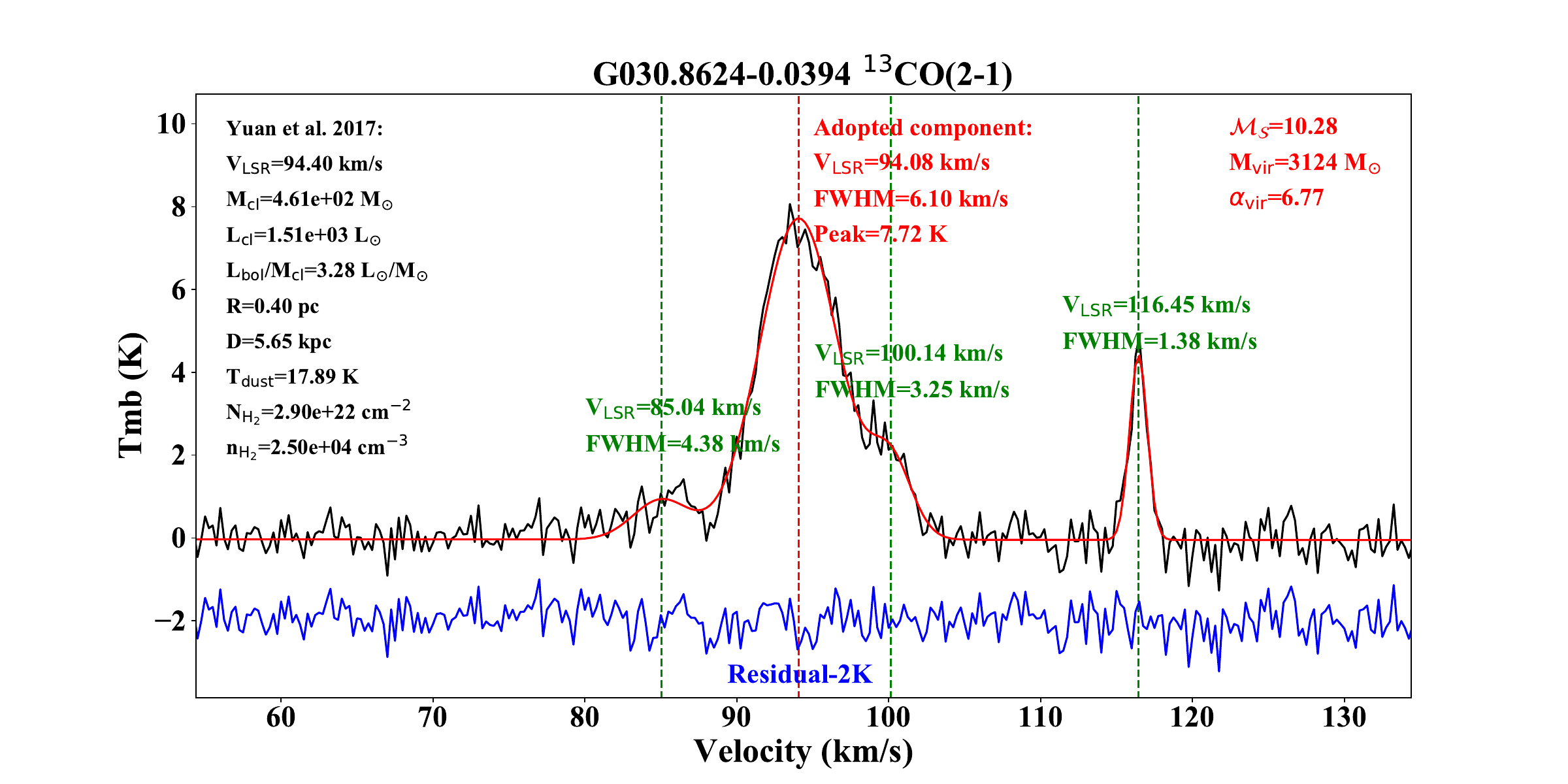}
	\caption{Example fitting results of HMSCs. The parameters in the left-top corner of the plots are from \citetalias{yuan2017high}. The red solid curve is the Gaussian fitting to the data. The red dashed line represents the mean value corresponding to the expected Gaussian component, and the red parameters next to the red dashed line represent the fitting results of the adopted component. The red parameters in the right-top corner are the Mach number, virial mass, and virial parameter of the source; the required B-field strength to support against gravity is also presented in the right-top corner if $\alpha_{\rm vir} <2$. The green dashed line and green parameters represent the center value of the other components and the fitting results of the corresponding components, respectively. The blue part is the residual shifted by 2 K for clarity.} 
	\label{fig:fit}
\end{figure*}

Figure~\ref{fig:fit} shows examples of the fitting results of HMSC sources.
The parameters in the left-top corner of the plot are from \citetalias{yuan2017high}, the estimation of these parameters is discussed in detail in Appendix~\ref{app:A}.
The fitting results of each clump are listed in Table~\ref{tab:parameter} in Appendix~\ref{app:D}, we then investigate the properties of HMSCs based on these results.

\subsection{Linewidth} \label{LW}
Assuming non-thermal motions in HMSCs are due to turbulence, we use linewidth to determine the turbulence of gas, the virial mass, and the virial parameter.
Here, we refer linewidth to the Full Width Half Maximum ($\Delta V_{\rm FWHM}$) of molecular emission. 
The linewidth values for these HMSCs are, in general, largely unavailable from the literature.
A small fraction of the \citetalias{yuan2017high}'s HMSCs have reported linewidth from different dense molecular tracers, but with quite different critical densities and mostly derived from poor spectral resolution (see Appendix~\ref{app:B} for details).
In order to better characterize the linewidths of these HMSCs, we use the SEDIGISM $^{13}$CO (2-1) line fitting results. The SEDIGISM $^{13}$CO (2-1) line data has sufficient spectral resolution to properly derive the linewidths, and also has a wide and uniform coverage for the HMSCs.

\begin{figure*}
	\centering
	\includegraphics[clip=true,trim=0cm 0cm 0cm 0.5cm,width=0.4\textwidth]{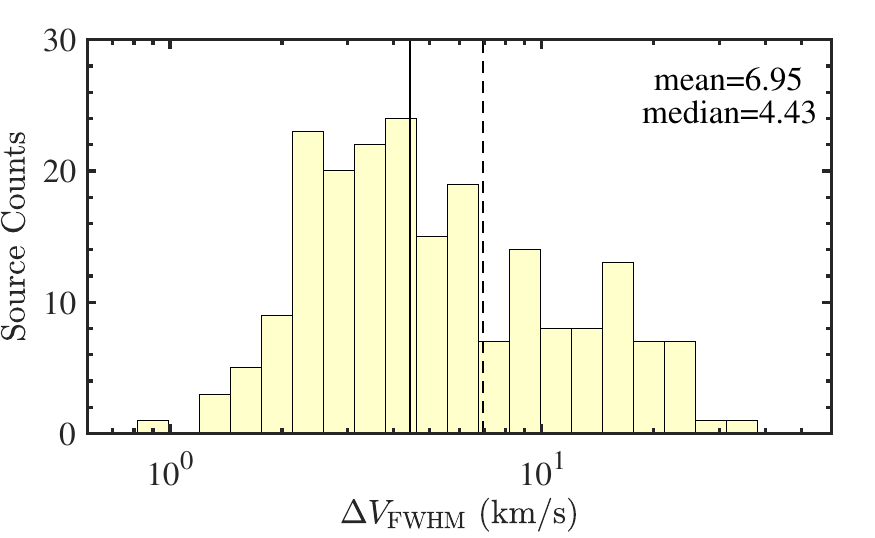}
	\includegraphics[clip=true,trim=0cm 0cm 0cm 0.5cm,width=0.4\textwidth]{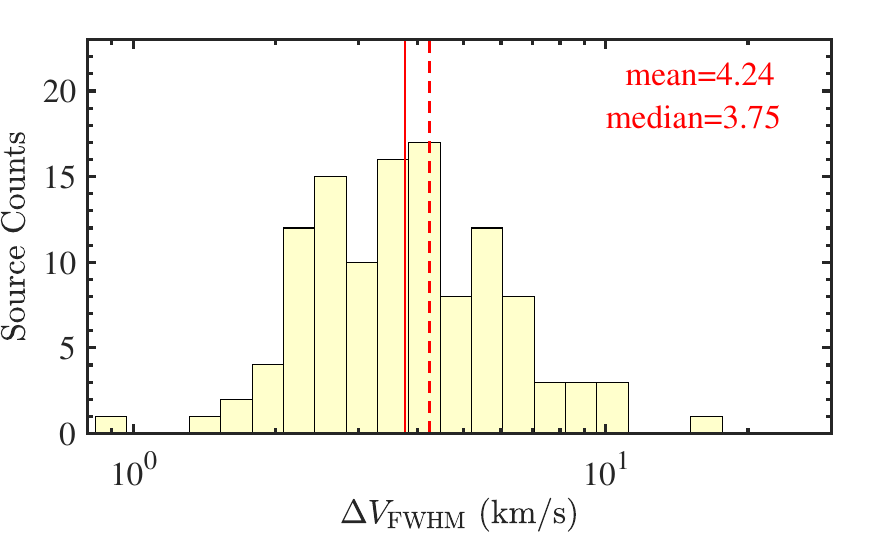}
	\includegraphics[clip=true,trim=0cm 0cm 0cm 0.5cm,width=0.4\textwidth]{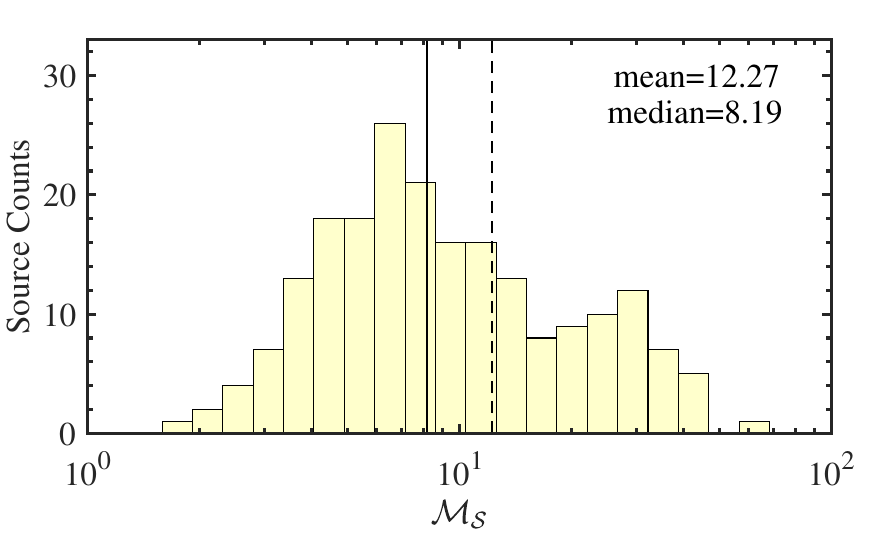}
	\includegraphics[clip=true,trim=0cm 0cm 0cm 0.5cm,width=0.4\textwidth]{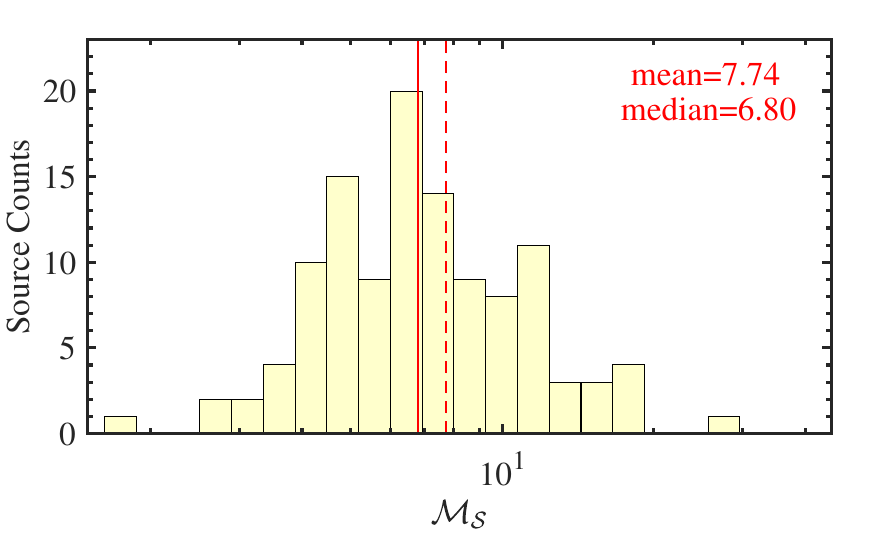}
	\includegraphics[clip=true,trim=0cm 0cm 0cm 0.5cm,width=0.4\textwidth]{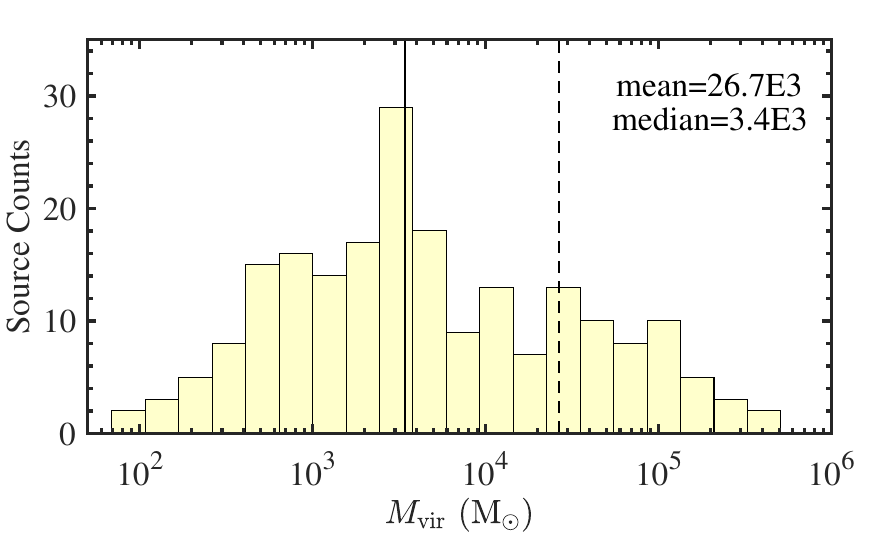}
    \includegraphics[clip=true,trim=0cm 0cm 0cm 0.5cm,width=0.4\textwidth]{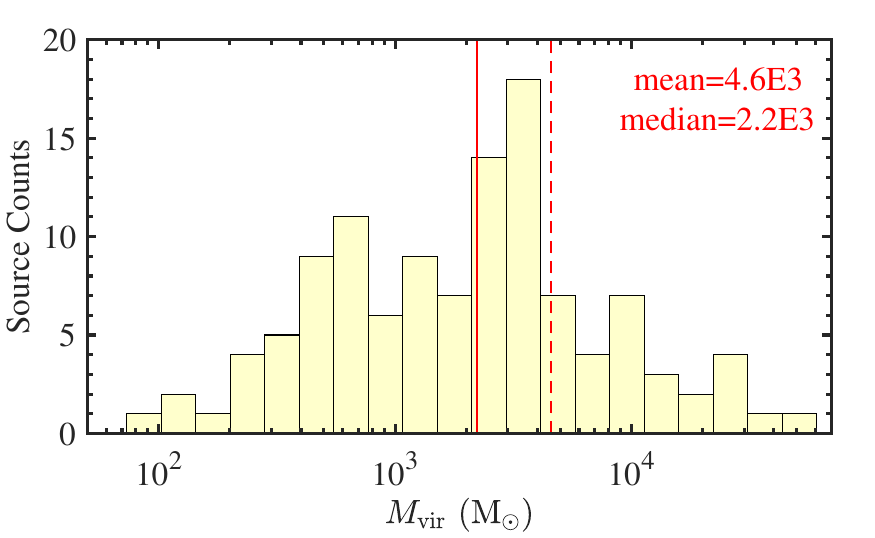}
	\includegraphics[clip=true,trim=0cm 0cm 0cm 0.5cm,width=0.4\textwidth]{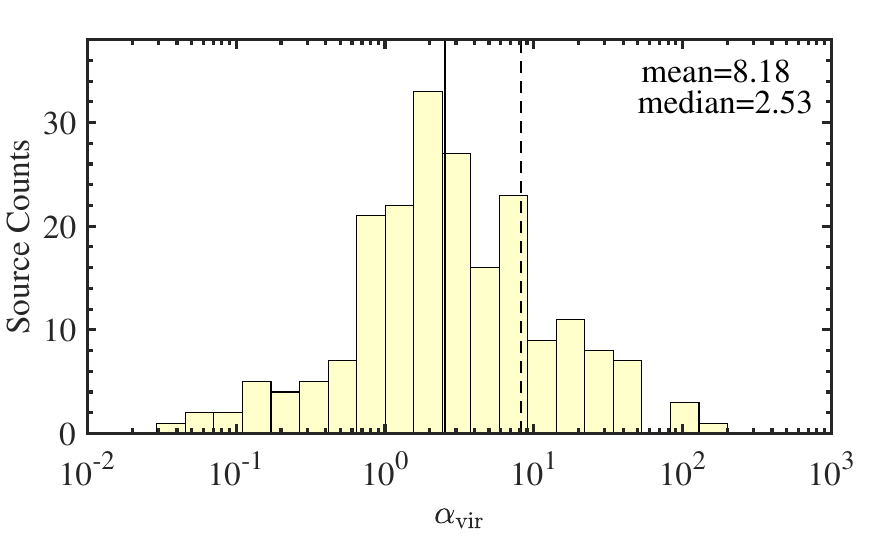}
	\includegraphics[clip=true,trim=0cm 0cm 0cm 0.5cm,width=0.4\textwidth]{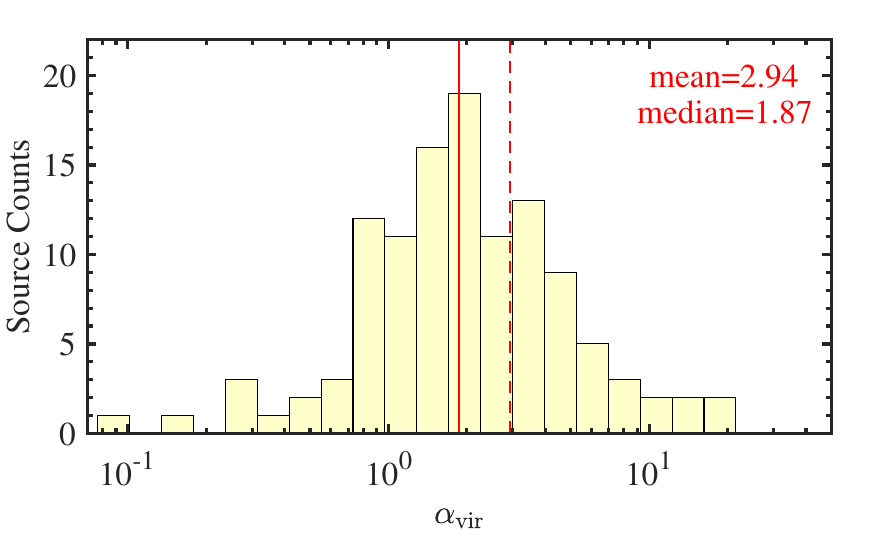}
	\includegraphics[clip=true,trim=0cm 0cm 0cm 0.5cm,width=0.4\textwidth]{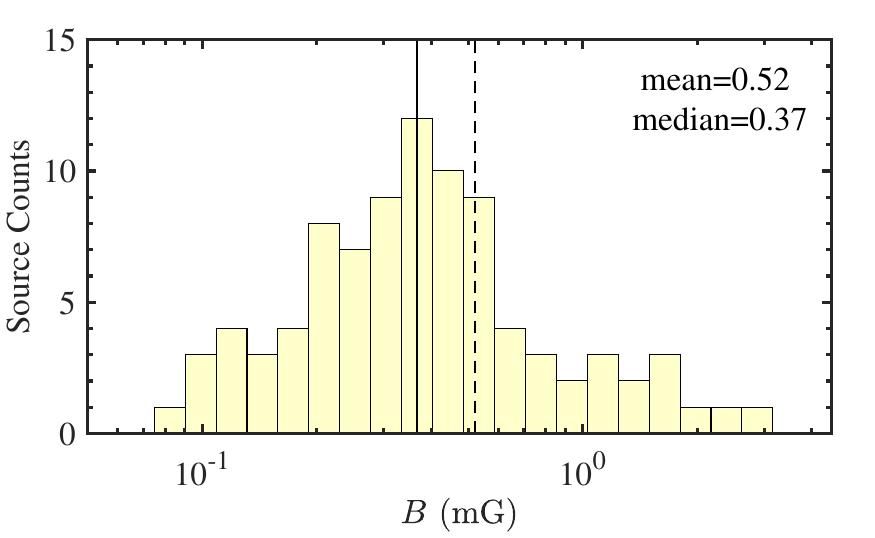}
	\includegraphics[clip=true,trim=0cm 0cm 0cm 0.5cm,width=0.4\textwidth]{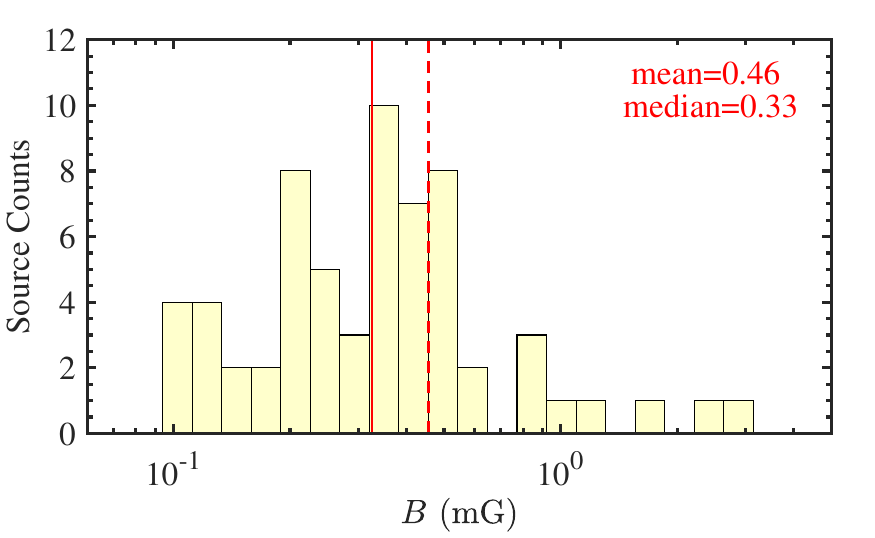}
	\caption{The histograms of Full Width Half Maximum $\Delta V_{\rm FWHM}$ (first row), Mach number $\mathcal{M_{S}}$ (second row), Virial mass $M_{\rm vir}$ (third row),  Virial parameter $\alpha_{\rm vir}$ (fourth row), and B-field $B$ (fifth row), respectively. The counting data of left-panels are from the full sample of HMSCs, while right-panels are from the sample excluding the sources located in $\vert l\vert \le 5^{\circ}$. Solid lines and dashed lines indicate the median value and mean value, respectively.} 
	\label{fig:hist}
\end{figure*}

The histogram of $\Delta V_{\rm FWHM}$ of HMSCs is shown in the first row of the left-side plot in Figure~\ref{fig:hist}, mean value and median value are 7.0 \kms and 4.4 \kms, respectively.
We then estimate the velocity dispersion by correcting the observed velocity dispersion, with the minor contribution of the velocity resolution:
\begin{eqnarray}\label{eq1}
\sigma_{\rm v}^{2}=\sigma_{\rm obs}^{2}-\frac{\Delta v_{\rm r}^{2}}{8\rm ln2}
\end{eqnarray}
where $\sigma_{\rm v}$ is the velocity dispersion along the line of sight, $\sigma_{\rm obs}=\Delta V_{\rm FWHM}/(2\sqrt{2\rm ln 2})$, and $\Delta v_{\rm r}$ is the velocity resolution of the observations $\sim 0.25\ \rm km\ s^{-1}$ \citep{schuller2017sedigism}.
The contribution of non-thermal motions to the velocity dispersion can be written as \citep{myers1983dense, fuller1992dense, sanchez2013properties, palau2015gravity, henshaw2016investigating, sokolov2018subsonic}:
\begin{eqnarray}\label{eq2}
\sigma_{\rm nth}^{2}=\sigma_{\rm v}^{2}-\sigma_{\rm th}^{2}=\sigma_{\rm v}^{2}-\frac{k_{\rm B}T_{\rm kin}}{m_{\rm obs}}
\end{eqnarray}
where $\sigma_{\rm nth}$ and $\sigma_{\rm th}$ are the non-thermal and the thermal velocity dispersion, respectively. $k_{\rm B}$ is the Boltzmann constant, $T_{\rm kin}$ is the kinetic temperature of the gas, $m_{\rm obs}$ is the mass of the observed molecule (for $^{13}$CO, $m_{\rm obs}=29\ m_{\rm H}$), and $m_{\rm H}$ is the mass of the hydrogen atom. 
Then we investigate the non-thermal components of the velocity dispersion with respect to the sound speed of the molecular hydrogen gas, which is referred to as the Mach number $\mathcal{M}_{S}$ \citep[e.g.,][]{palau2015gravity, wang2018temperature}:
\begin{eqnarray}\label{eq3}
\mathcal{M_{S}}=\frac{\sigma_{\rm nth}}{c_{s}}
\end{eqnarray}
here $c_{s}=\sqrt{(k_{\rm B}T_{\rm kin})/(\mu_{\rm m}m_{\rm H})}$ is the sound speed of the gas, and $\mu_{\rm m}=2.33$ is the mean molecular weight.

The second row of the left-side plot in Figure~\ref{fig:hist} shows the histogram of $\mathcal{M}_{S}$ by assuming $T_{\rm kin}=T_{\rm dust}$.
We find that most HMSCs are supersonic ($\mathcal{M}_{S}\gtrsim 5$), the range of Mach number is 1.7$\sim$62.1, with a mean and a median value of 12.3 and 8.2, respectively.
This result coincides with previous works on individual source.
For instance, \cite{carolan2009supersonic} find that cold massive core JCMT 18354-0649S is highly supersonic from analysis of both molecular line and dust continuum emission, and they suggest that the supersonic turbulence is mainly due to the large gravitational potential well at the center of the core.
However, \cite{sanhueza2017massive} find the Mach number of 1.1-1.8 toward a massive prestellar clump IRDC G028.23-00.19, which is transonic and mildly supersonic.
On the other hand, our results are in agreement with \cite{mckee2003formation}, which suggests that the cores that form massive stars are necessarily supersonically turbulent under the important premises of the turbulent core accretion model, but 22.2\% (46/207) of the clumps are mildly supersonic.
Note that the calculation of Mach number is in the one-dimensional case, it should be accounting for a factor of $\sqrt{3}$ when it converts to the three-dimensional: $(\mathcal{M_{S}})_{\rm 3D}=\sqrt{3}\mathcal{M_{S}}$ \citep[e.g.,][]{palau2015gravity}.


\subsection{Dynamical State} \label{DS}

For the reliable detection of 207 HMSCs with $^{13}$CO ($J$=2-1), we use line widths of the sources to calculate the virial mass $M_{\rm vir}$ and the virial parameter $\alpha_{\rm vir}$, which are further used to estimate the dynamic state of HMSCs.
The virial mass has been derived in \cite{maclaren1988corrections}, and it can be expressed as
\begin{eqnarray}\label{eq4}
M_{\rm vir}=3k\frac{R\sigma_{\rm v}^{2}}{G}
\end{eqnarray}
where $R$ is the equivalent radius of the clump, and $G$ is the gravitational constant.
$k=(5-2a)/(3-a)$ depend on the density profile $\rho\propto r^{-a}$ and it is discussed in \cite{maclaren1988corrections}, and the above expression can be rewritten as
\begin{eqnarray}\label{eq5}
M_{\rm vir}=210 \biggl( \frac{R}{\rm pc} \biggr) \biggl( \frac{ \Delta V_{\rm FWHM}}{\rm km \rm\ s^{-1}} \biggr)^{2} M_{\odot}
\end{eqnarray}

As shown in the third row of the left panel in Figure~\ref{fig:hist}, the mean and median value of $M_{\rm vir}$ are, 26700 $\rm M_{\odot}$ and 3400 $\rm M_{\odot}$, respectively.
Then the virial parameter \citep{bertoldi1992pressure} is defined as:
\begin{eqnarray}\label{eq6}
\alpha_{\rm vir}=\frac{M_{\rm vir}}{M_{\rm cl}}
\end{eqnarray}
which is used to evaluate the balance between gravity and the internal energy that can support the clump against gravitational collapse. Here $M_{\rm cl}$ is the clump mass.

Typically, some assumptions, such as the cloud is isolated or the surface terms are negligible, should be made when using virial equilibrium estimation, hence the virial parameter $\alpha_{\rm vir}$ is simply a measurement of the ratio between the kinetic term and gravitational energy.
However, clouds are interchanging mass, momentum, and energy with the surrounding medium, and surface energy terms are of the same order as the volume energy terms and can not be negligible \citep{ballesteros2006six, dib2007virial}.
Hence, converting a line width directly to a virial mass will overestimate it and thus makes it appear less bound than it actually is. 
According to \cite{kauffmann2013low}, $\alpha_{\rm vir}=1$ is gravitationally virial equilibrium and $\alpha_{\rm vir} \approx2$ is marginally gravitationally bound, while $\alpha_{\rm vir} \gtrsim2$ indicates an unbound clump.
The fourth row of the left panel in Figure~\ref{fig:hist} shows the histogram of $\alpha_{\rm vir}$, we can see that the median and mean value of $\alpha_{\rm vir}$ are 2.5 and 8.2, respectively.
We find that 43.5\% (90/207) of HMSC candidates have $\alpha_{\rm vir} \lesssim2$, implying that some HMSCs are gravitational bound and have the potential to further form stars, while more than half of the HMSCs have $\alpha_{\rm vir} \gtrsim2$ and are appearing to be unbound.

We do not take the magnetic field into account in the analysis so far, which could provide additional support against collapse and in turn increase the virial mass.
To assess the importance of the magnetic field in supporting against gravitational collapse, we follow \cite{henshaw2016investigating} and evaluate the virial parameter that includes the contribution of the magnetic field $\alpha_{\rm B,\rm vir}$, the formula is given by
\begin{eqnarray}\label{eq7}
\alpha_{\rm B,\rm vir}=\frac{5R}{GM_{\rm cl}}\biggl(\sigma_{\rm v}^{2}+\frac{\sigma_{\rm A}^{2}}{6}\biggr)
\end{eqnarray}
here $\sigma_{\rm A}=B(\mu_{0}\rho)^{-1/2}$ is the Alfvén velocity, in which $B$ is the magnetic field strength, $\mu_{0}$ is the permeability of free space, and $\rho$ is the density of the HMSCs.
We can estimate the minimum magnetic field strength to have a $\alpha_{\rm B,\rm vir}=2$ for the HMSCs with a $\alpha_{\rm vir}<2$. From Eq.\ref{eq7}, we obtain:
\begin{eqnarray}\label{eq8}
B=\sqrt{6\mu_{0}\rho \biggl(\frac{\alpha_{\rm B,\rm vir}GM_{\rm cl}}{5R}-\sigma_{\rm v}^{2}\biggr)}
\end{eqnarray}

The fifth row left plot in Figure~\ref{fig:hist} shows the strength of magnetic field required for the stability of bound clumps.
The range of B-field is 0.08$\sim$2.88 mG, with the median and mean value of 0.37 mG and 0.52 mG, respectively. 
For the observations towards high-mass starless cores or clumps, \cite{tan2013dynamics} present a $B$ value range of $0.09\sim 0.33$ mG, with a median value of 0.14 mG; \cite{tamaoki2019magnetic} find a total B-field strength of $0.11\sim 1.58$ mG, with a median value of 0.31 mG.
They are consistent with the derived results in this work.
Thus, for the typical HMSCs, the required B-field strength to support against gravity is in a reasonable range of observed values \citep{zhang2014magnetic}.


\section{Discussion} \label{Disc}

\begin{figure}
	\centering
	\includegraphics[clip=true,trim=0cm 0.25cm 0cm 0.25cm,width=0.55\textwidth]{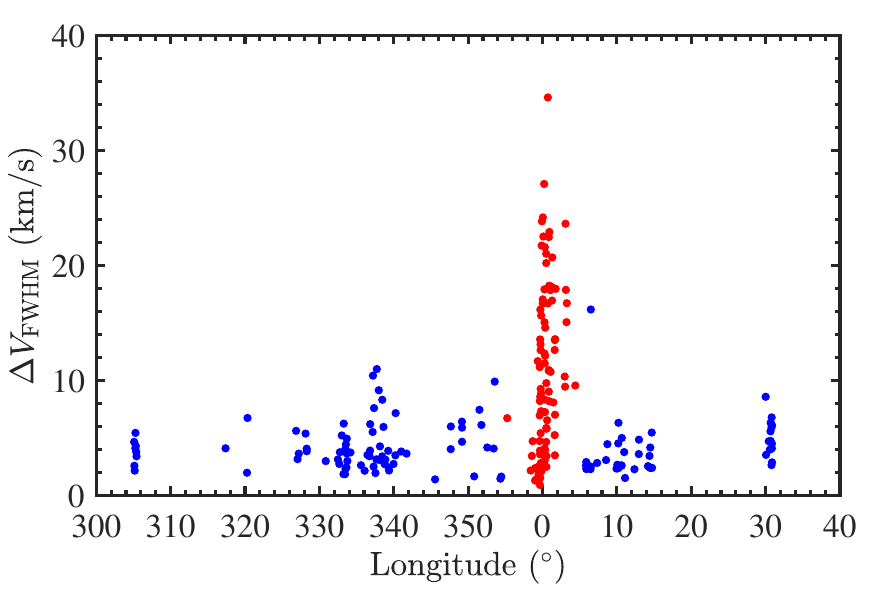}
	\caption{The distribution of $\Delta V_{\rm FWHM}$ in Galactic longitude toward 207 HMSC candidates. Red points indicate the HMSC sources within $\vert l\vert \le 5^{\circ}$, while blue points in the region of $\vert l\vert > 5^{\circ}$.} 
	\label{fig:distri_longitude}
\end{figure}

Most of the large velocity ($\geq$10 \kms) are from HMSCs around the galactic center. The HMSCs line width, $\Delta V_{\rm FWHM}$, has a wide range of $\sim$ 3 orders of magnitude (see Figure~ \ref{fig:distri_longitude}).
Some of the large linewidth may be affected by the complex kinematics in the Central Molecular Zone near the Galactic center.
In order to avoid the impact caused by it, we remove the sources near the galactic center, which are located in $\vert l\vert \le 5^{\circ}$ \citep[e.g.][]{yang2022sedigism}, and we investigate the rest sources to make a comparison with the whole sample.

After removing the 95 sources within $\vert l\vert \le 5^{\circ}$ (red points in Figure~\ref{fig:distri_longitude}), 112 HMSCs remain in the sample.
Row 1 to row 4 plots in the right-panel in Figure~\ref{fig:hist} show the histogram of this sample of $\Delta V_{\rm FWHM}$, Mach number $\mathcal{M_S}$, Virial mass $M_{\rm vir}$, and Virial parameter $\alpha_{\rm vir}$, respectively.
We can see that the median value of the parameters have a little difference comparing with the parameter distributions for sample including sources near the galactic center on the left side, with a difference of 15\% in $\Delta V_{\rm FWHM}$, 17\% in $\mathcal{M_{S}}$, 35\% in $M_{\rm vir}$, and 26\% in $\alpha_{\rm vir}$.
The fraction of bound clumps increases, with $\alpha\lesssim2$ is 55.4\% (62/112), comparing with the whole sample of 43.5\% (90/207).
The fifth row in the right-side plot in Figure~\ref{fig:hist} shows the histogram of B-field without the Galactic center sources.
We can see that the median value of B-field strengths required for the stability of bound clumps is 0.33 mG, which is also slightly different comparing to the full sample, with a difference of 11\%.
Thus, the distribution of HMSC candidates does have an impact on the linewidths, Mach numbers, virial masses, virial parameters, and also slightly affects the required B-field strengths to support HMSCs against collapse.
For convenience, the sources are arranged in three groups for the following discussions. Group \uppercase\expandafter{\romannumeral1}: the whole sample of 207 HMSCs, Group \uppercase\expandafter{\romannumeral2}: the 112 HMSCs in the region of $\vert l\vert > 5^{\circ}$, and Group \uppercase\expandafter{\romannumeral3}: the 95 HMSCs within $\vert l\vert \le 5^{\circ}$.

To better understand the HMSCs and the correlation between their dynamical properties, we combine our results with some parameters derived from \citetalias{yuan2017high}, and use Pearson correlation coefficient \citep{lee1988thirteen} to present the correlations between physical parameters for these sample of HMSCs, which is shown as percentages in Figure~\ref{fig:cor}.
The calculation is presented in Appendix~\ref{app:C}.
Here the data involved in the calculation of magnetic field correlation coefficient are derived from 90 (right panel: 62) bound clumps, others are from the total sample of 207 (right panel: 112) clumps.
We then discuss interesting correlation and relation between pairs of parameters: B-field and density, luminosity-to-mass ratio and dust temperature, as well as linewidth and size.

\begin{figure*}
	\centering
	\includegraphics[clip=true,trim=0.3cm 0.2cm 1.95cm 0cm,width=0.45
	\textwidth]{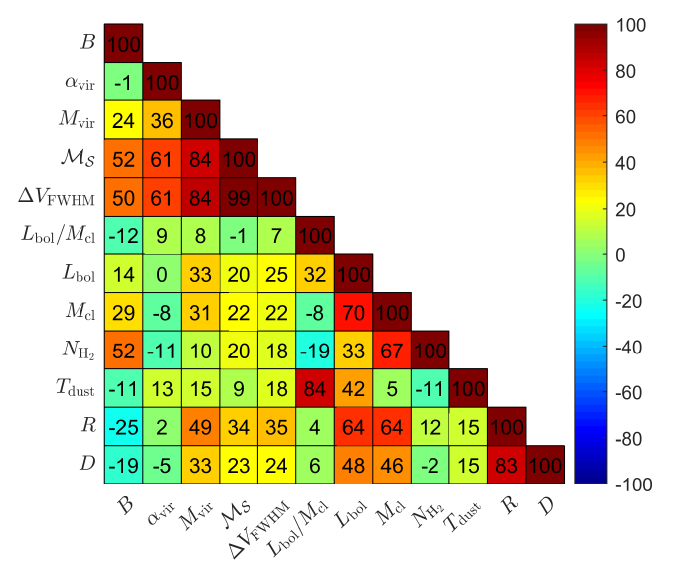}
 	\includegraphics[clip=true,trim=1.75cm 0.2cm 0.5cm 0cm,width=0.45
	\textwidth]{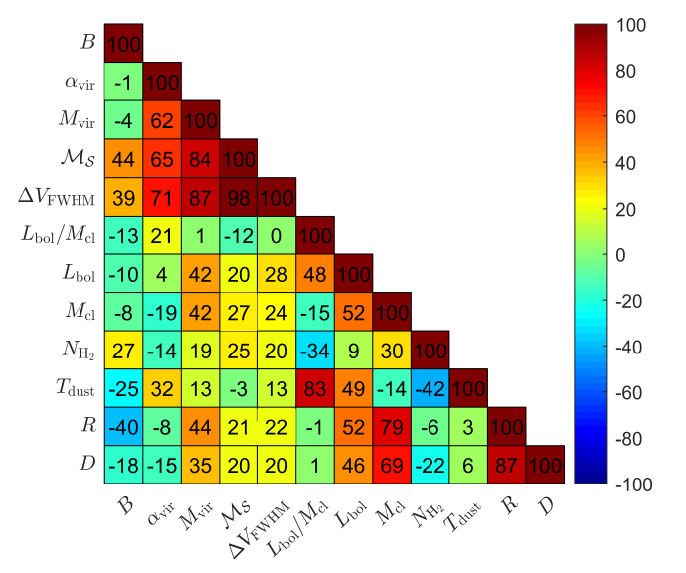}
	\caption{The correlation coefficient between the HMSCs parameters for Group \uppercase\expandafter{\romannumeral1} (left, whole sample of 207 sources) and Group \uppercase\expandafter{\romannumeral2} (right, 112 sources after excluding Galactic center), respectively. The correlation efficient have been counting a factor of 100. The range of $P$ is between -100 to 100. $P=0$ indicates that there is no relationship between the two sets of parameters, while $60\le \vert P \vert \le 100$ is considered to have a strong correlation. } 
	\label{fig:cor}
\end{figure*}

\begin{figure}
	\centering
	\includegraphics[clip=true,trim=0cm 0cm 0cm 0cm,width=0.48\textwidth]{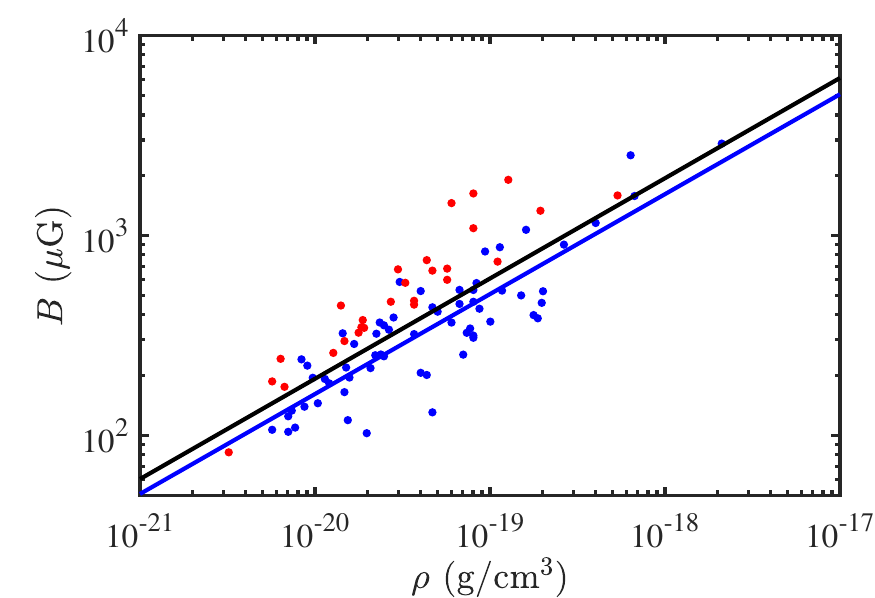}
	\includegraphics[clip=true,trim=0cm 0.0cm 0cm 0cm,width=0.48\textwidth]{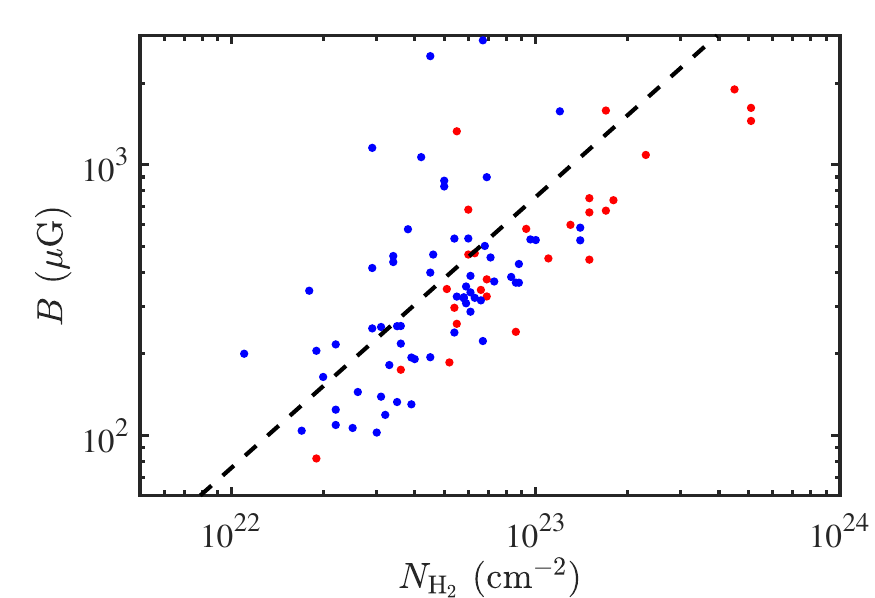}
	\caption{The scatter diagram between B-field and density (left), B-field and column density (right). Blue points indicate the HMSC sources in Group \uppercase\expandafter{\romannumeral2}, while red points in Group \uppercase\expandafter{\romannumeral3}. Black line and blue line are the fit for Group \uppercase\expandafter{\romannumeral1} and Group \uppercase\expandafter{\romannumeral2} points, respectively. The dashed line shows the mass-to-flux ratio $(M/\Phi)_{nor}=1$ from Eq.\ref{eq9}. The $(M/\Phi)_{nor}$ value have been normalized to the critical value, thus values greater than 1 are subcritical, and less than 1 are supercritical.} 
	\label{fig:B_rho}
\end{figure}

\subsection{Magnetic Field-Density Relation} \label{BDRelation}

Theory predicts that there is a relationship between gas density and B-field strength in high density regions, with a power law of $B\propto \rho ^\kappa$. 
We can find this relationship in Figure~\ref{fig:B_rho} and the B-field slightly increases with mass density, with $\kappa \approx$ 0.50 (for both Group \uppercase\expandafter{\romannumeral1} and Group \uppercase\expandafter{\romannumeral2}).
Keep in mind that the required B-field value to support against collapse is the upper limit for bound clumps, the B-field strength would be overestimated thus $\kappa$ would be less than 0.5.
For the two extreme-cases of B-field, both ambipolar diffusion (strong field) model and turbulence-driven (weak field) model predict that $\kappa \sim 0.5$ during the core collapse phase \citep{crutcher2004drives}, but B-field may also affect by the velocity dispersion $\sigma_{\rm v}$ if turbulence play a significant role in cloud support. 
It should be noted that \cite{crutcher1999magnetic} estimate for $\kappa$ is about 0.47, a result close to the theoretical prediction that the B-field would be strong enough to affect the cloud contraction, which is consistent with our results.

If the molecular cloud is well coupled to the B-field, the B-field in large-scale can provide support against gravitational collapse.
An important parameter in the discussion of the required B-field to resist the gravity in large scales is the mass-to-flux ratio $M/\Phi$, described as \citep{crutcher2004drives}:
\begin{eqnarray}\label{eq9}
\biggl(\frac{M}{\Phi} \biggr)_{nor}=7.6\times 10^{-21}\frac{N_{\rm H_{2}}}{B}\ \rm cm^{-2}\ \mu G^{-1}
\end{eqnarray}
where the subscript $nor$ indicates that $M/\Phi$ has been normalized by the critical value $(M/\Phi)_{cr}$.
Figure~\ref{fig:B_rho} right panel map shows the $B$-$N_{\rm H_{2}}$ distribution, black dashed line is the Eq.\ref{eq9} when $(M/\Phi)_{nor}=1$.
We can find from the plot that 27 (22 in Group \uppercase\expandafter{\romannumeral2}) points above dashed line are subcritical $(M/\Phi)_{nor}$, while 63 (40 in Group \uppercase\expandafter{\romannumeral2}) points lower dashed line are supercritical, with a fraction of 70.0\% (64.5\% in Group \uppercase\expandafter{\romannumeral2}).
The averaged mass-to-flux ratio in these HMSCs is $(M/\Phi)_{nor} \approx $1.32 cm$^{-2}$ $\mu$G$^{-1}$, which tends to be consistent with the turbulence-driven (i.e., weak magnetic field) model.
According to the result, we find that B-field support probably would not be strong enough to prevent collapse of self-gravitating HMSCs that are formed by compressible turbulence in such environments.

\subsection{Dust Temperature and Luminosity-to-Mass Ratio Correlation} \label{cor1}

\begin{figure*}
	\centering
	\includegraphics[clip=true,trim=0cm 0cm 0cm 0cm,width=0.48 \textwidth]{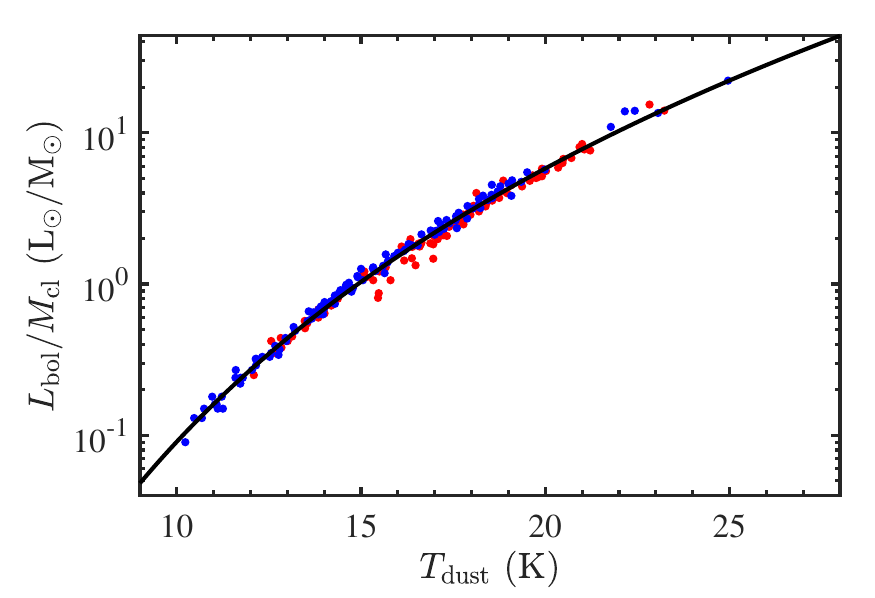}
	\includegraphics[clip=true,trim=0cm 0cm 0cm 0cm,width=0.48 \textwidth]{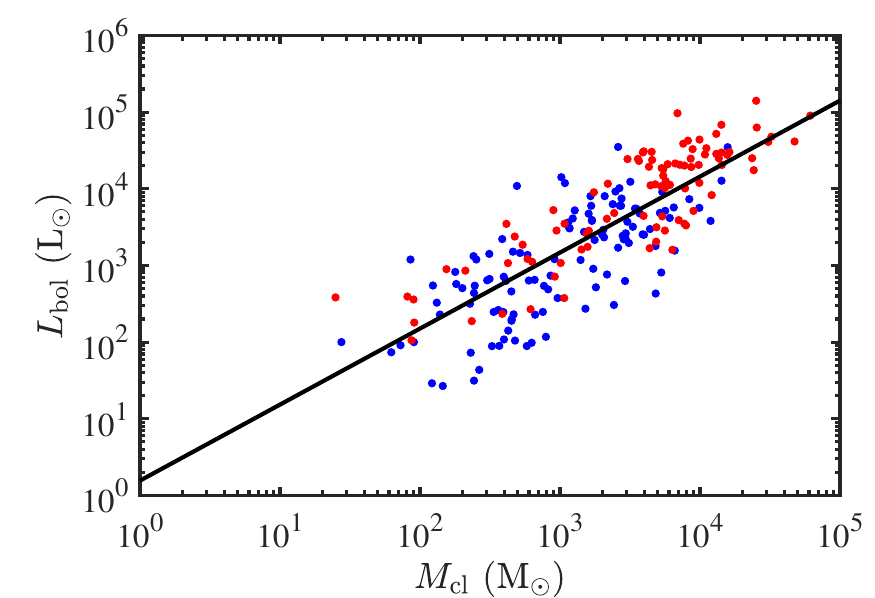}
	\caption{The scatter diagram between dust temperature and luminosity-mass ratio (left), bolometric luminosity and clump mass (right). Blue points indicate the HMSC sources in Group \uppercase\expandafter{\romannumeral2}, while red points in Group \uppercase\expandafter{\romannumeral3}. Black solid line is the power-law fit to the whole sample.} 
	\label{fig:fit_corr}
\end{figure*}

The bolometric luminosity ($L_{\rm bol}$) is the total power output of a source across all electromagnetic radiation wavelengths, and it is widely used in some astrophysics fields.
$T_{\rm dust}$ should have a strong correlation with $L_{\rm bol}$ if the $L_{\rm bol}$ arises from a greybody with a constant $T_{\rm dust}$ and assuming that dust emissivity index $\beta$ is fixed.
However, the measured $L_{\rm bol}$ values are usually highly uncertain, and they depend on the distance. 
Thus, the $L_{\rm bol}$-$T_{\rm dust}$ relation shows a looser correlation than that between the $L_{\rm bol}/M_{\rm cl}$ and $T_{\rm dust}$.
The luminosity-to-mass ratio $L_{\rm bol}/M_{\rm cl}$, a distance-independent parameter, is strongly correlated with $T_{\rm dust}$ as shown in the left panel in Figure~\ref{fig:fit_corr}. We find that the $(L_{\rm bol}/M_{\rm cl})$ strongly depends on the dust temperature.
The fitting result of the relation is:
\begin{eqnarray}\label{eq10}
L_{\rm bol}/M_{\rm cl}=(9.08\pm1.72)\times10^{-8} T_{\rm dust}^{6.00\pm 0.08}
\end{eqnarray}

\begin{figure}
	\centering
	\includegraphics[clip=true,trim=0cm 0cm 0cm 0cm,width=0.48\textwidth]{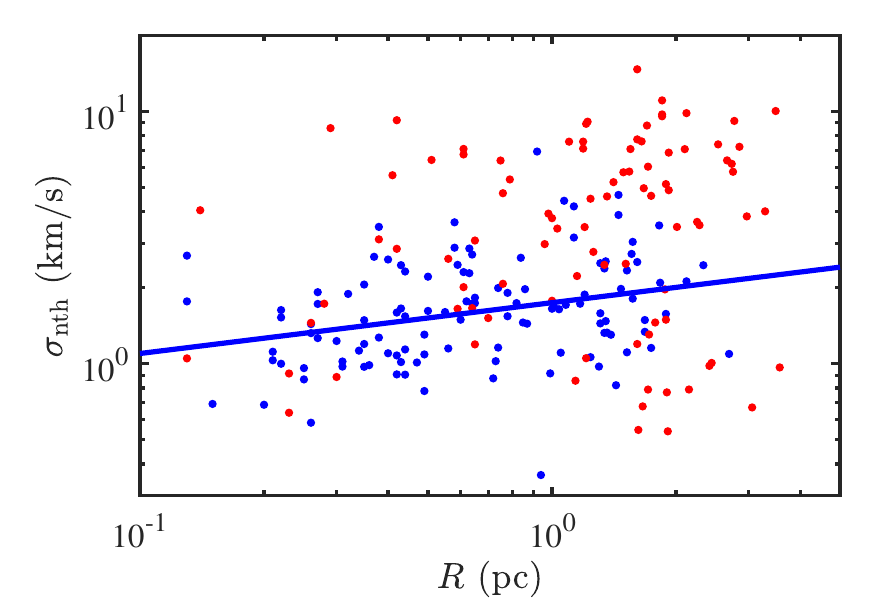}
	\includegraphics[clip=true,trim=0cm 0.0cm 0cm 0cm,width=0.48\textwidth]{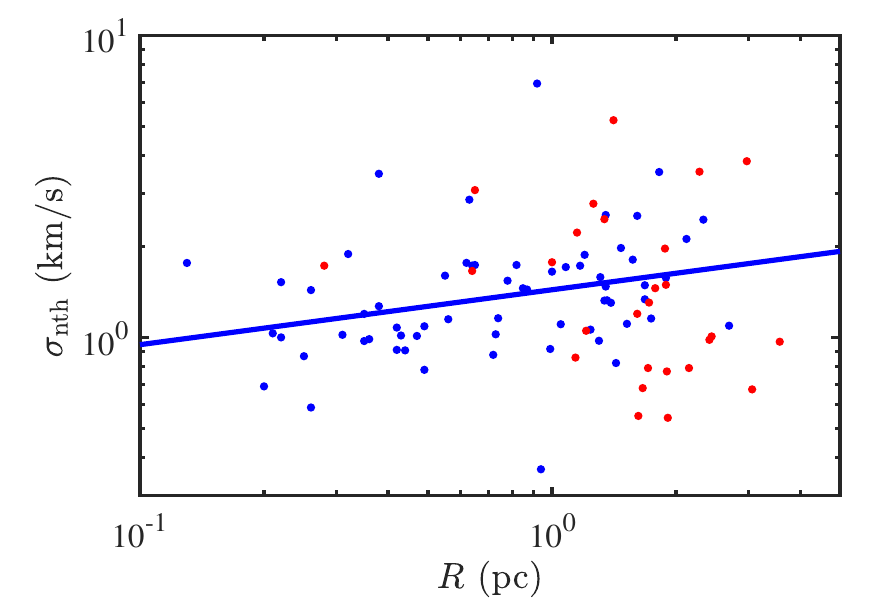}
	\caption{The scatter diagram between non-thermal velocity dispersion, based on the $^{13}$CO line, as a function of the equivalent radius, $R$, for the whole sample (left), and for the bound clumps (right). Blue points indicate the HMSCs in Group \uppercase\expandafter{\romannumeral2}, while red points in Group \uppercase\expandafter{\romannumeral3}. The blue lines show the power-law fit to the blue points.} 
	\label{fig:LWS}
\end{figure}

The correlation coefficient value is high, $P\approx 0.84$.
\cite{elia2017hi} and \cite{urquhart2018atlasgal} also found the similar result of the index within molecular clumps, with a power-law index of $\sim$ 6.00.
\cite{elia2016remarkable} used an analytic model for optically thin cold dust as a greybody, assuming a dust emissivity index $\beta$ of 2.0 \citep{ossenkopf1994dust} to derive the following relation:
\begin{eqnarray}\label{eq11}
L_{\rm bol}/M_{\rm cl}=1.71\times10^{-8} T_{\rm dust}^{6.00}
\end{eqnarray}
Within the uncertainties our results are in agreement with this relation.

Different phases of star formation have different luminosity-mass distribution, so $L_{\rm bol}$-$M$ is a useful parameter for separating different evolutionary stages of dense structures in molecular clouds  \citep[e.g.,][]{saraceno1996evolutionary, molinari2008evolution, traficante2015initial}.
The right panel in Figure~\ref{fig:fit_corr} shows the $L_{\rm bol}$-$M$ relation for the HMSCs, the fitting result is:
\begin{eqnarray}\label{eq12}
L_{\rm bol}\propto M_{\rm cl}^{0.99\pm 0.10}
\end{eqnarray}
The power-law index is only slightly lower than the value for  Class 0 sources, $\sim$ 1.13.
Observationally, it is much easier to measure dust temperature than luminosity-to-mass ratio, Eq.\ref{eq10} thus provides a possible and convenient way to use dust temperature to estimate luminosity-to-mass ratio, and to evaluate the evolutionary stage of the HMSCs.

\subsection{Linewidth-Size Relation} \label{LWSRelation}

\cite{larson1981turbulence} drew attention to the genuine turbulent nature of internal motion  to star forming regions based on the defining property of turbulent motion, and found a strong correlation between the linewidth and the size: molecular clouds are supersonically turbulent with linewidth that increase as a power of the size, i.e. $\Delta V_{\rm FWHM} \propto R^{\gamma}$.
He obtained the index $\gamma$ of this power-law relation is $\sim$ 0.38, similar to the Kolmogorov law (1/3) for turbulence in incompressible fluids, suggesting that the observed motions are all part of a common hierarchy of interstellar turbulent motions.
Later the linewidth-size scaling index was refined to 0.5 for Burgers turbulence \citep[e.g.,][]{solomon1987mass, passot1988plausibility}, a more appropriate model than Kolmogorov turbulence given the strongly supersonic conditions.
Some supersonic turbulent simulations, as well as analytical calculations of molecular clouds also support the idea of a linewidth-size relation with $\gamma=0.5$ \citep[e.g.,][]{vazquez1997search, padoan1999super, ballesteros2002physical}.
The power-law index of linewidth-size relation is thus important for developing a comprehensive understanding of interstellar medium, and used to study a common feature of interstellar turbulent motion for star-forming regions.
However, this relation seems to break in high-mass clumps and cores.
In these regions, the power-law index is in the 0.2--0.3 range \citep{caselli1995line, shirley2003cs}, significantly lower than that at larger scales.

\cite{traficante2018testing} and \cite{lu2022dynamical} found a very weak correlation between linewidth and size, with $\gamma=0.09$ and $\gamma=0.10$, respectively.
Moreover, some surveys did not find any clear relationship between linewidth and size for massive star-forming objects \citep{plume1997dense, wang2009relation, ballesteros2011gravity, kirk2017green, traficante2018massive}.

The average thermal term of velocity dispersion is $\sim$0.07 \kms\ for the HMSCs (for the expected temperature of 10-25 K), which is much less than that of the non-thermal component for most sources (see Table~\ref{tab:parameter}).
The left panel of Figure~\ref{fig:LWS} shows a weak correlation between the non-thermal velocity dispersion and the HMSCs radius for the sources of both Group \uppercase\expandafter{\romannumeral1} (correlation coefficient $P=0.35$), and Group \uppercase\expandafter{\romannumeral2} ($P=0.22$).
The right panel of Figure~\ref{fig:LWS} shows even a weaker correlation ($P=0.13$) for bound clumps in Group \uppercase\expandafter{\romannumeral1}.
The bound clumps in Group \uppercase\expandafter{\romannumeral2} show no significant difference with respect to the full sample of the same group.
To avoid the impact caused by Galactic center, we exclude Group \uppercase\expandafter{\romannumeral3}, and only perform power-law fit for the Group \uppercase\expandafter{\romannumeral2} sources, results are as follows:
\begin{eqnarray}\label{eq13}
\biggl(\frac{\sigma_{\rm nth}}{\rm km\ \rm s^{-1}}\biggr)\propto \biggl(\frac{R}{\rm pc}\biggr)^{0.20\pm0.12}
\end{eqnarray}
for all the clumps, and 
\begin{eqnarray}\label{eq14}
\biggl(\frac{\sigma_{\rm nth}}{\rm km\ \rm s^{-1}}\biggr)\propto \biggl(\frac{R}{\rm pc}\biggr)^{0.18\pm0.15}
\end{eqnarray}
for the bound clumps.

For the Group \uppercase\expandafter{\romannumeral2} clumps, the non-thermal velocity dispersion increases slowly with the equivalent radius for both, the whole sample and the bound massive clumps, but the correlation is weak.
The power-law indices found are similar to ones derived by \cite{caselli1995line} and \cite{lu2022dynamical}. Note that  \cite{lu2022dynamical} also found a weak correlation coefficient of $P=0.37$, consistent with our results. 
Massive star-forming regions are very dynamic, especially toward the densest regions, where star formation may take place, and which may be the regions of large scale converging flows. This makes difficult to retrieve the velocity scaling representative of the average turbulence cascade from the turbulent velocity field.
In addition, the uncertainty of the derived size should also be taken into account.
Therefore, previous and our studies show that the Larson scaling relation breaks at scales smaller than star-forming clumps of $\sim 1$\,pc. Use of the linewidth-size scaling relation to evaluate the turbulent motion for massive clumps should be taken more cautiously.

\section{Conclusion} \label{conclusions}
Following \citetalias{yuan2017high}, we have presented an analysis focusing on the global dynamical properties of a representative sample of 
207 HMSCs cross-matching the SEDIGISM and ATLASGAL surveys, and removing sources associated with outflows.
The sample has been selected in an unbiased way, covering a wide range of the Galactic plane ($-60^{\circ}<l<18^{\circ}$), and any known star formation signatures have been removed from the sample.
It is by far the most strictly selected sample, and should be regarded as a robust representative sample for HMSCs across our Galaxy. 
Our main results are summarized as follows:
\begin{enumerate}
    \item We decompose the thermal and non-thermal linewidth of $^{13}$CO (2-1) to evaluate the Mach number.
    We find that most HMSCs are supersonic, with a typical value of Mach number $\mathcal{M_{S}}\sim$ 8.2, suggesting that massive stars are forming in highly turbulent gas.
    
    \item The dynamical state of the HMSCs is evaluated by using the virial analysis, and we find that 44\%-55\% of HMSC are gravitationally bound, and the typical strength of magnetic field required to support against collapse of these HMSCs is 0.33$\sim$0.37 mG.
    
    \item Most of the HMSC sources are supercritical, with a mean value of $(M/\Phi)_{nor} \approx$ 1.32 cm$^{-2}$ $\mu$G$^{-1}$, which tends to be consistent with turbulence-driven (weak field) model, suggesting that the magnetic support would be insufficient to prevent collapse of self-gravitating bound HMSCs.
    
    \item The $L_{\rm bol}/M_{\rm cl}$ and $T_{\rm dust}$ have a strong correlation, with a power-law index of 6.00, suggesting that $L_{\rm bol}$ of the sources would be radiated from a greybody within a cold and dense environment.
    This strong correlation suggests that SED-derived dust temperature can be used as a proxy to evaluate the evolutionary stage of HMSCs.

    \item The linewidth-size scaling relation of the HMSCs have a power-law index of $\sim$0.2, but with a very weak correlation.
    One should be careful when considering linewidth-size relation in massive clumps, since the velocity scaling representative is difficult to retrieve in such dynamic and dense regions.   

\end{enumerate}

This study further advanced our understanding of global properties of HMSCs, and our high-resolution ALMA observations are ongoing to study the resolved properties.

\section*{Acknowledgements}
We acknowledge support from the National Science Foundation of China (11973013), the China Manned Space Project (CMS-CSST-2021-A09), the National Key Research and Development Program of China (2022YFA1603102), and the High-performance Computing Platform of Peking University.
This work is also partially supported by the program Unidad de Excelencia Mar\'ia de Maeztu CEX2020-001058-M.
BH and JMG acknowledge support by the grant PID2020-117710GB-I00 (MCI-AEI-FEDER, UE).
BH and EL acknowledge support from the National Natural Science Foundation of China (NSFC 12133003).
BH also acknowledges financial support from the China Scholarship Council (CSC) under grant No. 202006660008.
We acknowledge \'Alvaro S\'anchez-Monge for the useful discussions.
This publication is based on data acquired with the Atacama Pathfinder Experiment (APEX) under programmes 092.F-9315 and 193.C-0584.
APEX is a collaboration among the Max-Planck-Institut fur Radioastronomie, the European Southern Observatory, and the Onsala Space Observatory.
The processed data products are available from the SEDIGISM survey database located at \url{https://sedigism.mpifr-bonn.mpg.de/index.html}, which was constructed by James Urquhart and hosted by the Max Planck Institute for Radio Astronomy.


\bibliographystyle{aasjournal}


\appendix

\section{Calculation of Physical parameters of HMSCs} \label{app:A}

The distance to a source is a fundamental parameter that is essential to determine its mass and luminosity.
For all the HMSCs, distance estimates were obtained using a parallax-based distance estimator.
The equivalent radius $R$ is estimated by multiplying the distance $D$ by the deconvolved equivalent angular radius:
\begin{eqnarray}\label{ap01}
R=D\times\sqrt{\Theta_{\rm maj}\Theta_{\rm min}-\theta_{\rm HPBW}^{2}}
\end{eqnarray}
here $\theta_{\rm HPBW}=19\farcs2$ is the ATLASGAL beam.
The major and minor half-intensity axes ($\Theta_{\rm maj}$ and $\Theta_{\rm min}$) were obtained from \cite{csengeri2014atlasgal}.

The column density and the dust temperature are estimated via fitting far-infrared and submilimeter continuum data to the modified blackbody model. The model is expressed as:
\begin{eqnarray}\label{ap02}
I_{\nu}=B_{\nu}(T)(1-e^{-\tau_{\nu}})
\end{eqnarray}
where the Planck Function 
$B_{\nu}(T)=\biggl(\frac{2h \nu^{3}}{c^{2}}\biggr)\biggl(e^{\frac{h \nu}{k_{\rm B}T}}-1\biggr)^{-1}$ 
is modified by the optical depth:
\begin{eqnarray}\label{ap03}
\tau_{\nu}=\mu_{\rm H_{2}}m_{\rm H}\kappa_{\nu}N_{\rm H_{2}}/R_{\rm gd}
\end{eqnarray}
here $\mu_{\rm H_{2}}=2.8$ is the mean molecular weight, $m_{\rm H}$ is the mass of a hydrogen atom, $N_{\rm H_{2}}$ is the H$_{2}$ column density, $R_{\rm gd}=100$ is the gas-to-dust ratio.
The dust opacity $\kappa_{\nu}$ can be expressed as a power law in frequency:
\begin{eqnarray}\label{ap04}
\kappa_{\rm \nu}=3.33 \, \biggl(\frac{\nu}{600\ \rm GHz}\biggr)^{\beta}\ \rm cm^{2}\ \rm g^{-1}
\end{eqnarray}
$\beta$ is the dust emissivity index, and has been fixed to be 2 for the cold dust emission in the HMSCs.
The free parameters in this model are the dust temperature and the column density.
Then the fitting was performed by using python package \textit{lmfit} to estimate them.

The far-IR sources are taken from multiwavelength observations of Hi-GAL survey, with different angular resolutions (the effective angular resolutions are $10\farcs2$, $13\farcs5$, $18\farcs1$, $24\farcs9$, and $36\farcs4$ at 70, 160, 250, 350 and 500 $\mu$m, respectively).
The images were first convolved to a common angular resolution of $36\farcs4$, which is essentially the poorest resolution of any of the wavelengths.
Thus the clump mass is estimated by integrating the column densities in the scaled ellipses at a resolution of $36\farcs4$:
\begin{eqnarray}\label{ap05}
M_{\rm cl}=\mu_{\rm H_{2}} \, m_{\rm H}D^{2} \, \Omega_{\rm pix} \, \sum{N_{\rm H_{2}}}
\end{eqnarray}
here $d$ is the distance of the source, $\Omega_{\rm pix}$ is the solid angle of one pixel.
Under the angular resolution of $36\farcs4$, the major and minor axes of the scaled ellipses were
obtained by:
\begin{eqnarray}\label{ap06}
\Theta_{36.4}^{\rm maj}=\sqrt{(\Theta_{\rm atl}^{\rm maj})^2-19\farcs2^2+36\farcs4^2},\ \ \ \ \ 
\Theta_{36.4}^{\rm min}=\sqrt{(\Theta_{\rm atl}^{\rm min})^2-19\farcs2^2+36\farcs4^2}
\end{eqnarray}
$\Theta_{\rm atl}^{\rm maj}$ and $\Theta_{\rm atl}^{\rm min}$ are the major and minor axes given in the ATLASGAL catalog, respectively.

The H$_{2}$ number density of the HMSC candidates is calculated by:
\begin{eqnarray}\label{ap07}
n_{\rm H_{2}}=\frac{M_{\rm cl}}{(4/3)\pi R^{3}\mu_{\rm H_{2}}m_{\rm H}}
\end{eqnarray}

In the process of fitting the data, the frequency-integrated intensity $I_{\rm int}$ for each pixel is determined by using the resultant dust temperature and column density.
The luminosity of the sources with distance measurements were calculated by integrating $I_{\rm int}$ within the scaled ellipses:
\begin{eqnarray}\label{ap08}
L_{bol}=4\pi D^{2}\Omega_{\rm pix}\sum{I_{\rm int}}
\end{eqnarray}

\section{Line Widths Comparison} \label{app:B}

\begin{figure}
	\centering
	\includegraphics[clip=true,trim=1.3cm 4.9cm 1.4cm 6.1cm,width=0.42\textwidth]{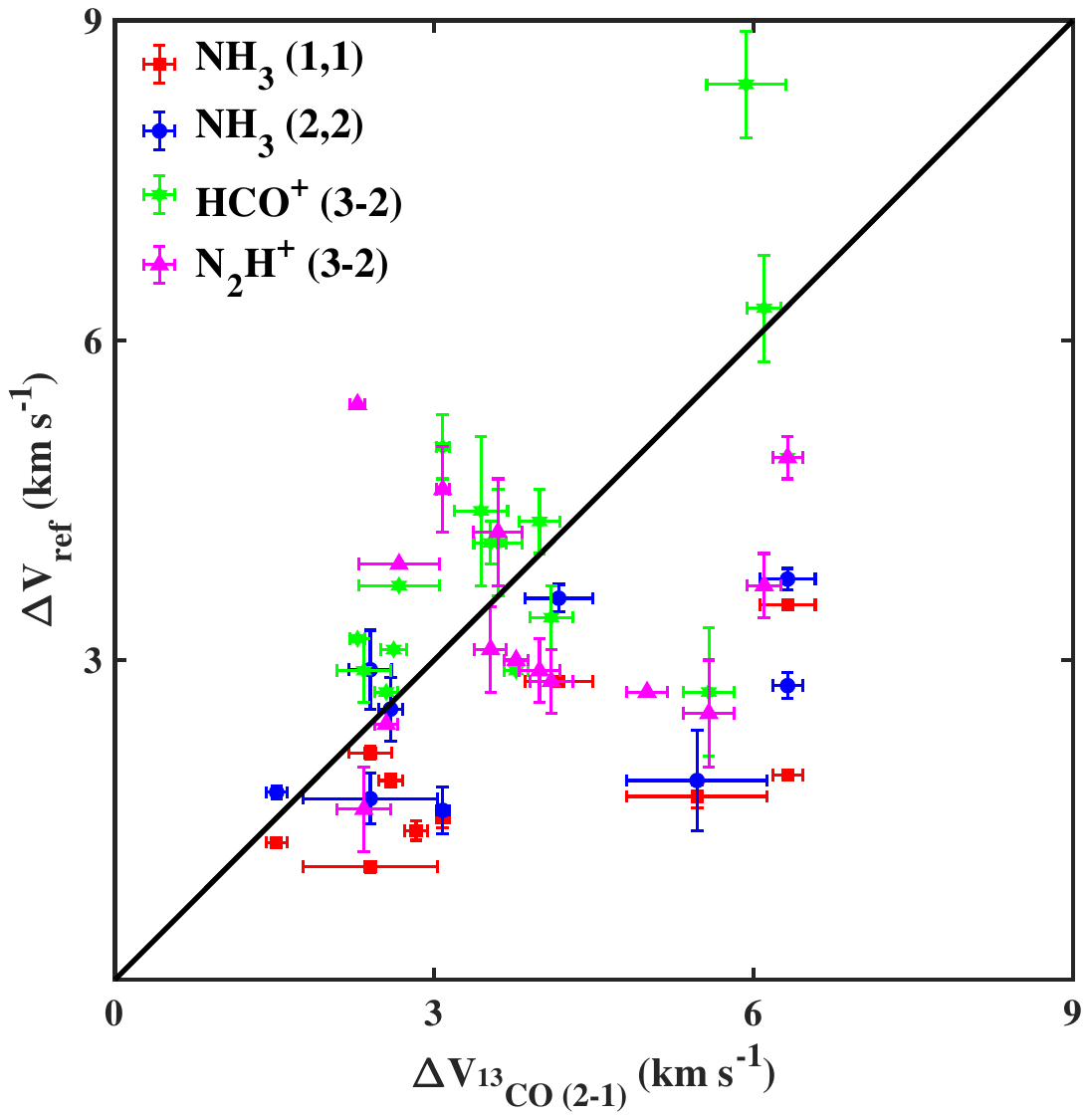}
	\caption{The difference of derived $\Delta V_{\rm FWHM}$ between $^{13}$CO (2-1) and other molecular lines reported by \cite{wienen2012ammonia}and \cite{shirley2013bolocam}. The black solid line represents when $\Delta V_{^{13}\rm CO (2-1)}=\Delta V_{\rm ref}$.} 
	\label{fig:con_FWHM}
\end{figure}

A small fraction of the HMSCs have linewidths reported in the literature, but these were obtained from different lines/transitions tracing  different critical densities, at different angular resolutions, and in most cases  with a poor spectral resolution. 
Owning to advantages in fine spectral resolution and a uniform and wide coverage of the Galaxy-wide HMSCs in question, in this paper we have adopted the $^{13}$CO (2-1) linewidths fitted from the SEDIGISM data.
Nevertheless, here we provide a summary of the literature linewidths and compare these with our $^{13}$CO (2-1) fitting results.

\cite{wienen2012ammonia} derived the NH$_{3}$ (1,1), (2,2), and (3,3) linewidths with a spectral resolution of 0.7 \kms; 10 sources of their catalog are also included in our sample.
\cite{shirley2013bolocam} also derived the linewidths from HCO$^{+}$ (3-2) and N$_{2}$H$^{+}$ (3-2) lines with a coarse spectral resolution of 1.1 \kms; 16 sources in our sample are covered.
Figure~\ref{fig:con_FWHM} shows a comparison of the linewidths reported in these works with the values derived from the $^{13}$CO (2-1) data.
We do not include NH$_{3}$ (3,3) because it traces hot gas heated by massive protostars.
The NH$_{3}$ linewidth are smaller than $^{13}$CO (2-1), but the rest are quite spread, with a tendency of having values smaller than the ones measured by $^{13}$CO (2-1).

Such differences may be due to several factors. An important one is the different critical densities.
For a typical HMSC dust temperature of $\sim$20 K, the critical density of the NH$_{3}$ (1,1), NH$_{3}$ (2,2), HCO$^{+}$ (3-2) and N$_{2}$H$^{+}$ (3-2) lines are about $2.0\times10^{3}$ cm$^{-3}$,  $1.6\times 10^{3}$ cm$^{-3}$, $1.6\times10^{6}$, $1.4\times10^{6}$, respectively.
Except for the ammonia lines, the $^{13}$CO (2-1) critical density,  $5.2\times10^{3}$ cm$^{-3}$, is smaller than these tracers \citep{shirley2015critical}.
The mean value of $n_{\rm H_{2}}$ in these HMSC candidates is $\sim 2.4 \times 10^{4}$ cm$^{-3}$. Therefore, the $^{13}$CO (2-1) line is a good tracer of HMSCs.
Another factor is the chemical evolution of the different molecules.
For example, ammonia is a late-type molecule and may appear only in more evolved regions of the HMSCs \citep[e.g.,][]{Taylor96}.

\section{Correlation between HMSC Physical Parameters} \label{app:C}

According to \cite{lee1988thirteen}, the correlation coefficient $P$ between parameter $X$ and parameter $Y$ is defined as
\begin{eqnarray}\label{ap09}
P(x,y)=\frac{\sum\limits_{i=1}^{n}(X_{x,i}-\overline X_{x})(Y_{y,i}-\overline Y_{y})}{\left[\sum\limits_{i=1}^{n}(X_{x,i}-\overline X_{x})^{2}\sum\limits_{j=1}^{n}(Y_{y,j}-\overline Y_{y})^{2}\right]^{\frac{1}{2}}}
\end{eqnarray}
where $n$ is the length of each column, $x$ and $y$ are the row number, $i$ and $j$ are the column number. $\overline X_{x}$ and $\overline Y_{y}$ are defined as the follow:
\begin{eqnarray}\label{ap10}
\overline X_{x}=\sum\limits_{i=1}^{n}\frac{(X_{x,i})}{n},\ \ \ \ \ \  \overline Y_{y}=\sum\limits_{j=1}^{n}\frac{(Y_{y,j})}{n}
\end{eqnarray}

The range of $P$ is between -1 to 1. $P=0$ indicates that there is no correlation between the two sets of parameters.
In general, $0.6\le \vert P \vert \le 1$ is considered to have a strong correlation.

\section{Physical Parameters of the HMSC Candidates}
\label{app:D}

\begin{longtable*}{| c || c | c | c | c | c | c | c | c | c | }
	\hline
        \multirow{2}*{Name} & $V_{\rm LSR}$ & $\Delta V_{\rm FWHM}$ & $\sigma_{\rm v}$ & $\sigma_{\rm th}$ & $\sigma_{\rm nth}$ & \multirow{2}*{$\mathcal{M}$} & $M_{\rm vir}$ & \multirow{2}*{$\alpha_{\rm vir}$} & $B$ \\
        ~ & ($\rm km\ s^{-1}$) & ($\rm km\ s^{-1}$) & ($\rm km\ s^{-1}$) & ($\rm km\ s^{-1}$) & ($\rm km\ s^{-1}$) & ~ & ($\rm M_{\odot}$) & ~ & (mG) \\
	\hline
	\endhead
	G000.0335+0.2051 & 111.8$\pm$0.8 & 17.1$\pm$1.3 & 7.24 & 0.07 & 7.24 & 28.95 & 1.7E+05 & 16.03 &  \\
	\hline
	G000.0355+0.2241 & 108.7$\pm$0.2 & 16.7$\pm$0.6 & 7.09 & 0.07 & 7.09 & 28.3 & 1.2E+05 & 22.49  &  \\
	\hline
	G000.0474+0.1124 & 102.9$\pm$0.3 & 23.2$\pm$0.7 & 9.84 & 0.07 & 9.84 & 38.96 & 2.4E+05 & 35.94  &  \\
	\hline
	G000.1150-0.1399 &  28.5$\pm$0.1 & 22.5$\pm$0.2 & 9.56 & 0.07 & 9.56 & 38.16 & 2.0E+05 & 25.46  &  \\
	\hline
	G000.2579+0.0182 &  40.8$\pm$0.2 &  8.3$\pm$0.6 & 3.54 & 0.07 & 3.54 & 14.44 & 3.3E+04 & 0.54  &  1.45 \\
	\hline
 	G000.2649+0.0356 &  27.8$\pm$0.7 &  12.4$\pm$1.2 & 5.24 & 0.07 & 5.24 & 21.70 & 4.5E+04 & 1.47  &  1.90 \\
	\hline
	G000.2670-0.0845 &  27.4$\pm$0.5 & 17.9$\pm$1.2 & 7.60 & 0.07 & 7.60 & 31.00 & 1.1E+05 & 19.55  &  \\
	\hline
	G000.2727-0.0636 &  36.3$\pm$0.2 & 15.1$\pm$0.5 & 6.39 & 0.07 & 6.39 & 24.28 & 1.3E+05 & 8.90   &  \\
	\hline
	G000.2836+0.0389 &  17.3$\pm$0.2 &  4.1$\pm$0.7 & 1.73 & 0.07 & 1.73 & 7.42 & 9.8E+02 & 0.97   &  1.58 \\
	\hline
	G000.2960+0.0438 &  15.1$\pm$0.8 & 11.5$\pm$2.3 & 4.87 & 0.07 & 4.87 & 20.09 & 5.3E+04 & 3.39   &  \\
	\hline
	G000.3069-0.1547 & 20.22$\pm$0.01 &  3.08$\pm$0.03 & 1.31 & 0.06 & 1.30 & 6.18 & 3.4E+03 &  0.43  &  0.33 \\
	\hline
	G000.3173+0.0117 &  25.3$\pm$0.1 & 21.6$\pm$0.2 & 9.17 & 0.08 & 9.17 & 31.96 & 2.7E+05 & 39.27  &  \\
	\hline
 	G000.3322+0.0373 &  28.3$\pm$0.3 & 7.3$\pm$0.7 & 3.08 & 0.07 & 3.08 & 11.66 & 7.2E+03 & 1.58  &  1.33  \\
	\hline
	G000.3404+0.0562 &  -6.9$\pm$0.3 &  3.9$\pm$0.9 &  1.66 & 0.07 & 1.66 & 6.75 & 2.1E+03 &  0.85  &  0.74 \\
	\hline
	G000.3550-0.0884 & 106.4$\pm$1.0 & 12.1$\pm$2.4 & 5.15 & 0.07 & 5.15 & 19.88 & 5.8E+04 &  6.59  &  \\
	\hline	
	G000.3659-0.0833 & 105.4$\pm$0.2 & 14.6$\pm$0.7 & 6.20 & 0.07 & 6.20 & 23.91 & 1.2E+05 &  9.34  &  \\
	\hline
	G000.4940-0.1144 &  42.4$\pm$0.8 & 21.0$\pm$1.2 & 8.93 & 0.08 & 8.93 & 33.16 & 1.1E+05 & 30.66  &  \\
	\hline
 	G000.4942+0.0194 &  54.7$\pm$0.5 & 4.7$\pm$1.4 & 1.97 & 0.07 & 1.97 & 8.41 & 8.5E+03 &  0.18  &  1.62  \\
	\hline
	G000.5041+0.1404 &  12.4$\pm$0.6 & 20.2$\pm$1.2 & 8.58 	& 0.08 & 8.58 & 32.32 & 2.5E+04 & 161.04 &  \\
	\hline
	G000.5136+0.1252 &  14.6$\pm$0.7 &  5.9$\pm$1.8 & 2.49 	& 0.07 & 2.49 & 9.57 & 1.1E+04 &  2.53  &  \\
	\hline
	G000.5151-0.1057 &  30.4$\pm$0.2 &  21.4$\pm$0.5 & 9.10 & 0.08 & 9.10 & 33.80 & 1.2E+05 &  26.08  &  \\
	\hline
	G000.5184-0.6127 & 16.97$\pm$0.02 &  2.49$\pm$0.04 & 1.05 & 0.07 & 1.05 & 4.06 & 1.7E+02 &  2.08  &  \\
	\hline
 	G000.5351-0.6329 & 16.77$\pm$0.02 &  3.43$\pm$0.04 & 1.45 & 0.07 & 1.453 & 5.60 & 6.4E+02 &  3.04  &  \\
	\hline
	G000.5568-0.0289 &  43.7$\pm$0.2 &  5.8$\pm$0.7 & 2.46 & 0.07 & 2.46 & 9.98 & 9.5E+03 &  0.97  &  0.67 \\
	\hline
	G000.6255-0.0956 &  62.0$\pm$0.3 &  6.5$\pm$1.3 & 2.77 	& 0.07 & 2.77 & 11.19 & 1.1E+04 &  0.79  &  1.09 \\
	\hline
	G000.7193-0.0723 &  59.0$\pm$0.4 & 34.6$\pm$1.2 & 14.69 & 0.07 & 14.69 & 62.09 & 4.0E+05 & 17.16  &  \\
	\hline
 	G000.7352+0.0065 &  79.1$\pm$0.6 & 16.7$\pm$1.7 & 7.09 	& 0.07 & 7.09 & 27.77 & 9.1E+04 & 17.05  &  \\
	\hline
	G000.8462-0.0019 &  78.8$\pm$0.5 & 10.8$\pm$1.1 & 4.60 & 0.07 & 4.60 & 18.72 & 3.4E+04 &  6.36  &  \\
	\hline
 	G000.8467-0.0292 &  87.7$\pm$0.4 & 9.0$\pm$0.5 & 3.83 &	0.07 & 3.83 & 15.92 & 5.1E+04 &  1.57  &  0.45  \\
	\hline	
	G000.8574-0.0935 &  98.1$\pm$0.6 & 10.9$\pm$0.8 & 4.63 & 0.07 & 4.63 & 18.64 & 4.3E+04 &  9.06  &  \\
	\hline
	G000.8785-0.1929 &  71.0$\pm$0.7 &  8.2$\pm$2.0 & 3.48 & 0.07 & 3.48 & 14.33 & 1.7E+04 &  7.83  &  \\
	\hline
 	G000.9029-0.0258 &  83.3$\pm$0.1 & 18.2$\pm$0.3 & 7.74 	& 0.07 & 7.74 & 33.06 & 1.1E+05 & 11.36  &  \\
	\hline
	G000.9295+0.0041 &  80.1$\pm$0.2 & 22.9$\pm$0.4 & 9.73 & 0.07 & 9.73 & 40.17 & 2.0E+05 & 33.44  &  \\
	\hline
	G001.0652-0.1338 &  72.2$\pm$0.4 & 17.9$\pm$0.9 & 7.59 & 0.07 & 7.58 & 31.49 & 7.4E+04 & 45.84  &  \\
	\hline
 	G001.2106+0.1020 & 119.2$\pm$0.4 & 17.4$\pm$0.9 & 7.40 	& 0.07 & 7.40 & 32.10 & 1.6E+05 & 40.81  &  \\
	\hline
	G001.2866+0.2941 & 110.1$\pm$0.4 & 16.8$\pm$0.8 & 7.12 & 0.07 & 7.12 & 30.75 & 7.0E+04 & 44.63  &  \\
	\hline
	G001.3159-0.0466 &  88.6$\pm$0.3 & 20.7$\pm$0.7 & 8.79 & 0.07 & 8.79 & 37.31 & 1.5E+05 & 19.54  &  \\
	\hline
 	G001.4536+0.2001 & 151.6$\pm$0.6 &  8.1$\pm$2.0 & 3.43 	& 0.07 & 3.42 & 14.33 & 1.4E+04 & 9.11   &  \\
	\hline	
	G001.6359-0.0278 &  51.5$\pm$0.3 & 12.7$\pm$0.6 & 5.37 & 0.06 & 5.37 & 25.96 & 2.7E+04 & 4.18   &  \\
	\hline	
	G001.6387-0.1889 &  40.4$\pm$0.7 &  5.3$\pm$2.1 & 2.23 & 0.06 & 2.22 & 9.99 & 6.6E+03 & 1.36   &  0.45 \\
	\hline
	G001.6627-0.1744 &  41.2$\pm$0.2 & 13.5$\pm$0.5 & 5.74 & 0.06 & 5.74 & 26.26 & 5.7E+04 & 10.18  &  \\
	\hline
	G001.6662-0.1394 &  47.3$\pm$0.2 &  4.19$\pm$0.5 & 1.78 & 0.06 & 1.77 & 8.27 & 3.7E+03 &  0.76  &  0.60 \\
	\hline
	G001.6825-0.0911 &  48.8$\pm$0.5 &  7.0$\pm$2.0 & 2.98 & 0.06 & 2.98 & 13.96 & 9.9E+03 & 2.27   &  \\
	\hline
	G001.6971-0.1706 &  45.0$\pm$0.3 & 13.6$\pm$0.8 & 5.77 & 0.06 & 5.77 & 26.36 & 6.0E+04 & 8.49   &  \\
	\hline
	G001.7640-0.4096 & -34.9$\pm$0.4 & 15.9$\pm$1.0 & 6.75 & 0.06 & 6.75 & 31.92 & 3.2E+04 & 30.21  &  \\
	\hline
	G002.9836+0.0107 &  62.9$\pm$0.3 & 10.6$\pm$0.6 & 4.51 & 0.07 & 4.50 & 19.55 & 2.9E+04 & 20.58  &  \\
	\hline
	G003.0371-0.0582 &  84.7$\pm$0.5 &  9.5$\pm$1.3 & 4.02 & 0.06 & 4.02 & 18.14 & 6.2E+04 & 5.10   &  \\
	\hline
 	G003.0928+0.1680 & 151.6$\pm$0.2 & 23.6$\pm$0.4 & 10.03 & 0.06 & 10.03 & 44.77 & 4.1E+05 & 16.91  &  \\
	\hline
	G003.1447+0.4733 &  71.9$\pm$0.7 & 17.9$\pm$1.9 & 7.59 & 0.06 & 7.59 & 33.74 & 8.0E+04 & 87.00  &  \\
	\hline
 	G003.2278+0.4924 & 16.8$\pm$0.4 & 15.1$\pm$1.3 & 6.42 &	0.06 & 6.42 & 28.46 & 2.4E+04 & 104.50 &  \\
	\hline
	G003.2702+0.4446 &  26.9$\pm$0.3 & 16.7$\pm$1.0 & 7.10 & 0.06 & 7.10 & 32.07 & 3.6E+04 & 92.49  &  \\
	\hline	
 	G004.4076+0.0993 &  9.8$\pm$0.2 &  9.6$\pm$0.4 & 4.06 & 0.07 & 4.05 & 16.00 & 2.7E+03 & 29.84 & \\
	\hline	
	G005.8523-0.2397 &  16.82$\pm$0.04 &  2.6$\pm$0.1 & 1.10 & 0.06 & 1.10 & 5.30  & 5.7E+02 &  2.47  &  \\
	\hline
	G005.8799-0.3591 & 6.25$\pm$0.04  &  2.9$\pm$0.1 & 1.23 & 0.07 & 1.23 & 5.08  & 5.3E+02 &  2.19  &  \\
	\hline
	G005.8917-0.3567 & 5.93$\pm$0.02  &  2.3$\pm$0.1 & 0.97 & 0.07 & 0.97 & 3.96  & 3.9E+02 &  1.59  &  0.25 \\
	\hline
	G005.9394-0.3754 &  6.30$\pm$0.03  & 2.3$\pm$0.1 & 0.98 & 0.07 & 0.97 & 3.95  & 3.5E+02 &  2.64  &  \\
	\hline
	G007.3449-0.0001 &  19.7$\pm$0.1 & 2.8$\pm$0.1  & 1.20 & 0.06 & 1.20 & 5.87  & 5.9E+02 &  1.60  &  0.37 \\
	\hline
	G008.5441-0.3421 &  36.49$\pm$0.02 &  3.1$\pm$0.1 & 1.30 & 0.06 & 1.30 & 6.24  & 9.8E+02 &  2.29  &  \\
	\hline
        G008.7264-0.3959 &  39.65$\pm$0.04 &  4.5$\pm$0.1 & 1.89 & 0.06 & 1.89 & 8.97  & 1.3E+03 &  1.77  &  0.90 \\
	\hline
	G009.9517-0.3649 &  13.2$\pm$0.1 &  2.3$\pm$0.3 & 0.99 & 0.06 & 0.99 & 4.78 & 4.1E+02 &  1.27  &  0.31 \\
	\hline	
	G010.0676-0.4076 &  11.3$\pm$0.1 &  2.7$\pm$0.4 & 1.13 & 0.07 & 1.13 & 4.36  & 5.1E+02 &  4.11  &  \\
	\hline
	G010.1839-0.4050 &  15.7$\pm$0.1 &  4.5$\pm$0.2 & 1.92 & 0.08 & 1.92 & 6.81  & 1.2E+03 & 13.61  &  \\
	\hline
	G010.1976-0.2876 & 10.15$\pm$0.04 &  2.4$\pm$0.1 & 1.00 & 0.07 & 1.00 & 3.95  & 2.6E+02 &  1.43  & 0.46 \\
	\hline
        G010.2074-0.3051 & 11.4$\pm$0.1 &  6.3$\pm$0.1 & 2.68 & 0.07 & 2.68 & 11.04 & 1.1E+03 &  3.47  &  \\
	\hline	
	G010.6174-0.0304 &  63.4$\pm$0.1 &  2.6$\pm$0.1 & 1.11 & 0.07 & 1.10 & 4.60  & 1.5E+03 &  1.02  &  0.16 \\
	\hline
	G010.6233-0.5091 &  -2.5$\pm$0.1 &  2.6$\pm$0.1 & 1.09 & 0.06 & 1.09 & 4.82  & 3.8E+03 &  0.27  &  0.24 \\
	\hline
	G010.6627+0.0811 &  19.7$\pm$0.1 &  5.0$\pm$0.2 & 2.12 & 0.06 & 2.12 & 10.21 & 1.1E+04 &  0.93  &  0.32 \\
	\hline
	G010.9823-0.3677 &  -0.6$\pm$0.1 &  3.7$\pm$0.1 & 1.58 & 0.06 & 1.57 & 6.96  & 5.5E+03 &  0.97  &  0.19 \\
	\hline
	G011.0970-0.1093 &  30.8$\pm$0.2 &  2.4$\pm$0.4 & 1.02 & 0.06 & 1.06 & 5.03  & 3.8E+02 &  0.96  &  0.50 \\
	\hline
        G012.3628+0.4214 &  31.40$\pm$0.03 &  2.3$\pm$0.1 & 0.96 & 0.07 & 0.96 & 4.18  & 2.7E+02 &  3.02  &  \\
	\hline
	G012.9459-0.2488 &  33.4$\pm$0.1 &  3.6$\pm$0.2 & 1.53 & 0.06 & 1.52 & 7.71  & 6.0E+02 &  0.95  &  1.57 \\
	\hline
	G012.9674-0.2380 &  33.8$\pm$0.1 &  4.9$\pm$0.2 & 2.06 & 0.06 & 2.06 & 10.10 & 1.7E+03 &  3.63  &  \\
	\hline
	G014.1842-0.2280 & 39.94$\pm$0.04 &  2.6$\pm$0.1 & 1.08 & 0.06 & 1.08 & 4.98  & 5.7E+02 &  1.23  &  0.33 \\
	\hline
	G014.3848-0.1308 &  23.5$\pm$0.1 &  3.4$\pm$0.3 & 1.46 & 0.05 & 1.45 & 7.64  & 2.1E+03 &  0.44  &  0.83 \\
	\hline
	G014.4683-0.0857 &  37.6$\pm$0.1 &  2.4$\pm$0.2 & 1.01 & 0.06 & 1.01 & 4.55 & 5.7E+02 &  0.74  &  0.43 \\
	\hline
        G014.4876-0.1274 &  40.8$\pm$0.3 &  4.2$\pm$0.3 & 1.77 & 0.06 & 1.76 &  7.97 & 4.7E+02 &  1.20  &  2.88 \\
	\hline
	G014.6858-0.2234 &  39.3$\pm$0.3 &  5.5$\pm$0.7 & 2.32 & 0.06 & 2.32 & 10.82 & 2.8E+03 &  6.13  &  \\
	\hline
	G014.7258-0.2031 &  37.8$\pm$0.3 &  2.4$\pm$0.6 & 1.02 & 0.06 & 1.01 & 5.20  & 5.2E+03 &  0.66  &  0.53 \\
	\hline
	G030.0062-0.1192 &  99.2$\pm$0.1 &  8.6$\pm$0.3 & 3.63 & 0.06 & 3.63 & 16.52 & 8.9E+03 & 10.84  &  \\
	\hline
	G030.0556+0.0995 &  97.2$\pm$0.1 &  3.5$\pm$0.1 & 1.49 & 0.06 & 1.49 & 6.57  & 1.6E+03 &  2.38  &  \\
	\hline
	G030.4235-0.2142 & 104.37$\pm$0.03 &  4.7$\pm$0.1 & 2.00 & 0.07 & 1.99 & 7.99  & 3.4E+03 &  2.95  &  \\
	\hline
	G030.5682-0.0258 &  88.8$\pm$0.1 &  3.9$\pm$0.3 & 1.66 & 0.07 & 1.65 & 7.17  & 1.4E+03 &  2.30  &  \\
	\hline
	G030.6574+0.0446 &  81.6$\pm$0.2 &  5.8$\pm$0.3 & 2.46 & 0.06 & 2.46 & 10.81 & 4.2E+03 &  9.30  &  \\
	\hline
	G030.6783-0.0386 &  86.8$\pm$0.1 &  6.4$\pm$0.3 & 2.71 & 0.07 & 2.71 & 10.95 & 5.5E+03 &  3.24 &  \\
	\hline
        G030.6858-0.0306 & 90.6$\pm$0.2 &  16.3$\pm$0.3 & 6.93 & 0.07 & 6.93 & 27.55 & 5.1E+04 &  18.70\\
	\hline
	G030.7912-0.1173 &  94.8$\pm$0.3 &  5.8$\pm$0.5 & 2.46 & 0.07 & 2.46 & 10.35 & 3.0E+03 &  7.37  &  \\
	\hline	
	G030.7941+0.0736 &  36.7$\pm$0.1 &  2.6$\pm$0.2 & 1.12 & 0.07 & 1.11 & 4.74  & 3.1E+02 &  4.93  &  \\
	\hline
	G030.8130-0.0235 &  94.95$\pm$0.03 & 6.8$\pm$0.1 & 2.88 & 0.07 & 2.88 & 11.08 & 5.6E+03 &  2.11  &  \\
	\hline
	G030.8447+0.1775 &  95.6$\pm$0.1 &  4.1$\pm$0.2 & 1.73 & 0.06 & 1.73 & 8.78  & 2.2E+03 &  1.48  &  0.45 \\
	\hline
	G030.8523-0.1086 &  100.0$\pm$0.2&  4.5$\pm$1.2 & 1.91 & 0.07 & 1.91 & 7.75  & 3.3E+03 &  2.83  &  \\
	\hline
        G030.8620+0.0392 &  93.75$\pm$0.04 &  2.7$\pm$0.1 & 1.14 & 0.07 & 1.14 & 4.35  & 6.7E+02 &  2.66  &  \\
	\hline
	G030.8624-0.0394 &  94.1$\pm$0.1 &  6.1$\pm$0.2 & 2.59 & 0.07 & 2.59 & 10.28 & 3.1E+03 &  6.77  &  \\
	\hline	
	G305.0943+0.2510 & -38.85$\pm$0.03 &  4.7$\pm$0.1 & 1.97 & 0.07 & 1.97 & 8.40  & 3.9E+03 &  4.28  &  \\
	\hline
	G305.1293-0.0271 & -38.26$\pm$0.03 &  2.6$\pm$0.1 & 1.09 & 0.07 & 1.09 & 4.51  & 6.8E+02 &  1.72  &  0.21 \\
	\hline
	G305.1721+0.0079 & -31.51$\pm$0.02 &  2.16$\pm$0.04 & 0.91 & 0.08 & 0.91 & 3.05  & 4.1E+02 &  0.83  &  0.34 \\
	\hline	
	G305.2581+0.3275 & -39.4$\pm$0.1 &  4.1$\pm$0.1 & 1.74 & 0.07 & 1.74 & 7.44  & 2.9E+03 &  1.64  &  0.32 \\
	\hline
	G305.2719-0.0309 & -32.93$\pm$0.02 &  5.4$\pm$0.1 & 2.31 & 0.07 & 2.31 & 9.09  & 3.8E+03 &  3.07  &  \\
	\hline
	G305.3187+0.3130 & -40.01$\pm$0.02 &  4.3$\pm$0.1 & 1.83 & 0.08 & 1.83 & 6.86  & 2.5E+03 &  6.55  &  \\
	\hline
        G305.3834+0.2565 & -35.7$\pm$0.1 &  3.8$\pm$0.1 & 1.60 & 0.07 & 1.60 & 6.30  & 1.7E+03 &  1.45  &  0.47 \\
	\hline
	G305.4126+0.2061 & -38.9$\pm$0.2 &  3.4$\pm$0.5 & 1.44 & 0.06 & 1.43 & 6.75  & 6.3E+02 &  0.95  &  1.15 \\
	\hline	
        G317.3887+0.1195 & -41.4$\pm$0.1 &  4.1$\pm$0.1 & 1.74 & 0.07 & 1.74 & 7.46  & 2.3E+03 &  1.15  &  0.58 \\
	\hline
	G320.2715+0.2920 & -31.42$\pm$0.02 &  0.9$\pm$0.1 & 0.37 & 0.06 & 0.36 & 1.74  & 1.6E+02 &  0.09  &  0.25 \\
	\hline
	G320.3385-0.1534 & -6.5$\pm$0.1  &  6.7$\pm$0.2 & 2.86 & 0.07 & 2.86 & 12.01 & 6.0E+03 &  1.75  &  1.07 \\
	\hline
        G326.8693+0.0720 & -49.1$\pm$0.5 &  5.6$\pm$0.6 & 2.38 & 0.06 & 2.38 & 10.43 & 8.9E+03 &  3.01  &  \\
	\hline
	G327.0678-0.2901 & -58.9$\pm$0.2 &  3.2$\pm$0.3 & 1.34 & 0.06 & 1.34 & 6.01  & 3.5E+03 &  1.14  &  0.13 \\
	\hline
	G327.2309-0.5041 & -49.01$\pm$0.03 &  3.6$\pm$0.1 & 1.54 & 0.07 & 1.54 & 6.27  & 1.2E+03 &  4.04  &  \\
	\hline
        G328.1434-0.1092 & -89.4$\pm$0.1 &  5.4$\pm$0.2 & 2.28 & 0.06 & 2.28 & 10.11 & 3.8E+03 &  4.48  &  \\
	\hline
	G328.2972-0.5130 & -44.8$\pm$0.2 &  4.1$\pm$0.6 & 1.73 & 0.06 & 1.72 & 7.66  & 9.4E+02 &  2.80  &  \\
	\hline
        G328.2988-0.5242 & -45.0$\pm$0.3 &  3.9$\pm$0.7 & 1.63 & 0.07 & 1.63 & 6.91 & 6.8E+02 &  3.01  &  \\
	\hline
	G330.8521-0.3510 & -61.90$\pm$0.01 &  2.99$\pm$0.04 & 1.26 & 0.07 & 1.26 & 4.80  & 5.1E+02 &  2.10  &  \\
	\hline
	G332.4947-0.1215 & -49.80$\pm$0.02 &  3.13$\pm$0.04 & 1.33 & 0.07 & 1.32 & 5.30  & 2.8E+03 &  1.72  &  0.11 \\
	\hline
	G332.6826-0.0082 & -96.8$\pm$0.1 &  2.7$\pm$0.3 & 1.16 & 0.05 & 1.16 & 6.01  & 1.2E+03 &  0.48  &  0.53 \\
	\hline
	G332.7570-0.4666 & -52.89$\pm$0.01 &  3.76$\pm$0.03 & 1.60 & 0.07 & 1.59 & 6.14  & 1.2E+03 &  7.00  &  \\
	\hline
	G333.0151-0.4964 & -54.10$\pm$0.02 &  5.2$\pm$0.1 & 2.21 & 0.07 & 2.21 & 8.86  & 2.9E+03 &  4.87  &  \\
	\hline
	G333.2418-0.5148 & -49.3$\pm$0.1 &  1.9$\pm$0.1 & 0.78 & 0.07 & 0.78 & 3.14  & 3.6E+02 &  1.77  &  0.10 \\
	\hline
	G333.2770-0.4846 & -55.1$\pm$0.1 &  6.3$\pm$0.2 & 2.65 & 0.06 & 2.65 & 13.28 & 3.0E+03 &  5.24  &  \\
	\hline
	G333.4489-0.1822 & -42.3$\pm$0.1 &  2.1$\pm$0.2 & 0.88 & 0.07 & 0.87 & 3.75  & 6.5E+02 &  0.18  &  0.87 \\
	\hline
	G333.4974-0.1154 & -48.64$\pm$0.04 &  4.0$\pm$0.1 & 1.71 & 0.07 & 1.71 & 6.90  & 3.7E+03 &  1.55  &  0.25 \\
	\hline
	G333.5081-0.2398 & -50.3$\pm$0.1 &  2.3$\pm$0.1 & 0.98 & 0.07 & 0.97 & 4.06  & 1.5E+03 &  0.27  &  0.39 \\
	\hline
	G333.5685+0.0292 & -85.4$\pm$0.3 &  4.4$\pm$0.5 & 1.88 & 0.06 & 1.88 & 8.50  & 4.9E+03 &  1.24  &  0.34 \\
	\hline
        G333.6425+0.3764 & -35.12$\pm$0.03 &  2.4$\pm$0.1 & 1.03 & 0.06 & 1.03 & 5.17  & 2.6E+02 &  1.81  &  0.40 \\
	\hline
	G333.6708-0.3575 & -48.78$\pm$0.02 &  3.5$\pm$0.1 & 1.49 & 0.07 & 1.49 & 6.21  & 9.1E+02 &  6.53  &  \\
	\hline
	G333.6827-0.2555 & -47.52$\pm$0.03 &  4.9$\pm$0.1 & 2.09 & 0.08 & 2.09 & 7.32  & 9.4E+03 &  3.60  &  \\
	\hline
        G333.7656+0.3477 & -33.59$\pm$0.03 &  3.0$\pm$0.1 & 1.27 & 0.06 & 1.27 & 5.70  & 7.2E+02 &  1.98  &  0.32 \\
	\hline
	G334.1747+0.0764 & -40.74$\pm$0.04 &  3.7$\pm$0.1 & 1.58 & 0.07 & 1.58 & 6.71  & 3.8E+03 &  1.89  &  0.14 \\
	\hline
	G335.5915+0.1836 & -49.22$\pm$0.02 &  2.6$\pm$0.1 & 1.11 & 0.06 & 1.11 & 5.05  & 2.2E+03 &  0.85  &  0.14 \\
	\hline
	G336.0895+0.0341 & -117.68$\pm$0.04&  2.2$\pm$0.1 & 0.91 & 0.07 & 0.90 & 3.53  & 4.3E+02 &  1.37  &  0.20 \\
	\hline
	G336.4689-0.2023 & -24.4$\pm$0.2 &  3.5$\pm$0.4 & 1.49 & 0.06 & 1.49 & 6.64  & 4.4E+03 &  1.53  &  0.12 \\
	\hline
        G336.7428+0.1078 & -77.1$\pm$0.1 &  3.4$\pm$0.1 & 1.45 & 0.08 & 1.44 & 5.15  & 3.2E+03 &  3.1  &  \\
	\hline
	G336.8177+0.1268 & -82.3$\pm$0.1 &  3.9$\pm$0.2 & 1.65 & 0.07 & 1.65 & 6.48  & 3.2E+03 &  1.53  &  0.25 \\
	\hline	
	G336.8315+0.1307 & -80.1$\pm$0.1 &  6.2$\pm$0.1 & 2.63 & 0.07 & 2.63 & 10.28 & 6.8E+03 &  4.05  &  \\
	\hline
	G337.1617+0.1113 & -71.5$\pm$0.1 &  5.5$\pm$0.5 & 2.34 & 0.06 & 2.34 & 10.51 & 9.7E+03 &  3.31  &  \\
	\hline
	G337.2059-0.0742 & -68.0$\pm$0.2 &  10.4$\pm$0.4 & 4.42 & 0.07 & 4.42 & 18.77 & 2.4E+04 &  6.94  &  \\
	\hline
	G337.2860+0.0083 & -105.2$\pm$0.2&  2.5$\pm$0.8 & 1.06 & 0.06 & 1.06 & 5.24 & 1.6E+03 &  0.25  &  0.53 \\
	\hline
        G337.3478-0.1584 & -70.9$\pm$0.2 &  8.2$\pm$0.4 & 3.48 & 0.06 & 3.48 & 17.10 & 5.4E+03 &  1.84  &  2.52 \\
	\hline
	G337.5371-0.1728 & -55.3$\pm$0.1 &  2.0$\pm$0.2 & 0.82 & 0.07 & 0.82 & 3.5  & 1.1E+03 &  0.38  &  0.18 \\
	\hline
	G337.6682-0.0430 & -47.5$\pm$0.2 &  3.1$\pm$0.7 & 1.33 & 0.06 & 1.33 & 5.84  & 2.8E+03 &  0.85  &  0.22 \\
	\hline
	G337.7207+0.0436 & -50.1$\pm$0.4 & 11.0$\pm$0.8 & 4.66 & 0.07 & 4.66 & 18.20 & 3.7E+04 &  14.76 &  \\
	\hline
        G337.9882+0.0237 & -45.9$\pm$0.4 &  9.2$\pm$0.9 & 3.88 & 0.06 & 3.88 & 17.20 & 2.6E+04 &  9.09  &  \\
	\hline
	G338.0911-0.1493 & -44.3$\pm$0.2 &  3.1$\pm$0.5 & 1.30 & 0.07 & 1.30 & 5.00  & 2.8E+03 &  1.68  &  0.10 \\
	\hline
	G338.1464+0.1094 & -37.9$\pm$0.2 &  4.3$\pm$0.3 & 1.81 & 0.06 & 1.81 & 8.21  & 6.0E+03 &  1.54  &  0.19 \\
	\hline
	G338.4460-0.0064 & -47.9$\pm$0.5 &  3.4$\pm$1.1 & 1.44 & 0.07 & 1.44 & 5.87 & 2.1E+03 &  0.78  &  0.44 \\
	\hline
        G338.4598+0.0239 & -40.13$\pm$0.04 &  8.3$\pm$0.1 & 3.53 & 0.07 & 3.53 & 14.34 & 2.6E+04 &  1.68  &  0.59 \\
	\hline
	G338.6209+0.0228 & -23.2$\pm$0.1 &  6.0$\pm$0.2 & 2.53 & 0.06 & 2.53 & 11.35 & 1.2E+04 &  1.98  &  0.29 \\
	\hline
	G338.7851+0.4767 & -65.2$\pm$0.1 &  2.7$\pm$0.3 & 1.15 & 0.06 & 1.15 & 5.42  & 8.7E+02 &  0.90  &  0.37 \\
	\hline
	G338.8688-0.4796 & -36.7$\pm$0.1 &  3.1$\pm$0.1 & 1.32 & 0.06 & 1.32 & 6.79  & 5.3E+02 &  2.19  &  \\
	\hline
	G339.2580-0.0069 & -106.6$\pm$0.1&  3.9$\pm$0.2 & 1.64 & 0.06 & 1.64 & 7.30  & 3.3E+03 &  2.34  &  \\
	\hline
        G339.3148+0.2296 & -69.9$\pm$0.1 &  2.4$\pm$0.2 & 1.02 & 0.07 & 1.02 & 4.10  & 9.0E+02 &  1.74  &  0.12 \\
	\hline
	G339.3645-0.1812 & -45.13$\pm$0.04 &  2.2$\pm$0.1 & 0.92 & 0.06 & 0.91 & 4.23  & 9.8E+02 &  0.57  &  0.22 \\
	\hline	
	G339.9677-0.1632 & -36.12$\pm$0.04 &  2.7$\pm$0.1 & 1.16 & 0.07 & 1.15 & 4.70  & 2.7E+03 &  1.02  &  0.11 \\
	\hline
	G340.2230-0.1681 & -52.26$\pm$0.04 &  3.48$\pm$0.1 & 1.47 & 0.06 & 1.47 & 6.93  & 3.4E+03 &  0.71  &  0.32 \\
	\hline
        G340.2545-0.2261 & -46.9$\pm$0.1 &  7.2$\pm$0.2 & 3.04 & 0.06 & 3.04 & 13.45 & 1.7E+04 &  2.02  &  \\
	\hline
	G341.0096-0.3595 & -45.7$\pm$0.1 &  3.8$\pm$0.5 & 1.62 & 0.06 & 1.62 & 7.55 & 1.5E+03 &  3.39  &  \\
	\hline
	G341.7206+0.0591 & -51.6$\pm$0.1 &  3.6$\pm$0.1 & 1.54 & 0.07 & 1.54 & 6.71  & 2.2E+02 &  1.05  &  0.42 \\
	\hline
        G345.5554+0.0239 & -17.4$\pm$0.1 &  1.4$\pm$0.1 & 0.59 & 0.07 & 0.58 & 2.53  & 1.1E+02 &  1.48  &  0.13 \\
	\hline
	G347.6622+0.2264 & -93.19$\pm$0.04 &  4.0$\pm$0.1 & 1.69 & 0.08 & 1.69 & 6.08  & 3.4E+03 &  3.11  &  \\
	\hline	
	G347.6783+0.2039 & -73.4$\pm$0.4 &  6.0$\pm$0.7 & 2.54 & 0.06 & 2.54 & 12.83 & 1.0E+04 &  1.93  &  0.35 \\
	\hline
	G349.1665+0.1308 & -64.8$\pm$0.4 &  6.4$\pm$0.4 & 2.72 & 0.06 & 2.72 & 11.90 & 1.3E+04 &  2.61  &  \\
	\hline
	G349.1770+0.1247 & -62.3$\pm$0.4 &  4.7$\pm$0.6 & 1.98 & 0.06 & 1.98 & 8.73  & 6.7E+03 &  1.03  &  0.37 \\
	\hline
	G349.1823+0.0645 & -73.5$\pm$0.1 &  5.9$\pm$0.3 & 2.50 & 0.06 & 2.50 & 11.29 & 9.6E+03 &  2.18  &  \\
	\hline
	G350.8162+0.5146 & -0.62$\pm$0.03  &  1.7$\pm$0.1 & 0.70 & 0.07 & 0.69 & 2.73  & 8.7E+01 &  3.15  &  \\
	\hline
        G351.5228+0.1963 & -42.7$\pm$0.1 &  7.5$\pm$0.3 & 3.16 & 0.07 & 3.16 & 12.34 & 1.3E+04 &  4.14  &  \\
	\hline
	G351.7858+0.2130 & -41.8$\pm$0.2 &  5.8$\pm$0.5 & 2.46 & 0.06 & 2.46 & 11.21 & 1.6E+04 &  1.66  &  0.22 \\
	\hline
	G352.5730-0.1914 & -83.98$\pm$0.04 &  4.2$\pm$0.1 & 1.77 & 0.07 & 1.77 & 7.22  & 2.3E+03 &  1.34  &  0.53 \\
	\hline
        G353.4298-0.1939 & -82.5$\pm$0.1 &  4.1$\pm$0.2 & 1.73 & 0.06 & 1.73 & 8.19  & 4.1E+03 &  1.89  &  0.19 \\
	\hline
	G353.5734-0.0798 & -56.3$\pm$0.3 &  9.9$\pm$0.7 & 4.20 & 0.07 & 4.20 & 16.32 & 2.3E+04 & 18.24  &  \\
	\hline	
	G354.3403+0.4742 &  3.9$\pm$0.2  &  2.1$\pm$0.3 & 0.87 & 0.06 & 0.87 & 4.37  & 2.2E+02 &  0.84  &  0.53 \\
	\hline
	G354.4525+0.4192 &  4.7$\pm$0.1  &  1.6$\pm$0.2 & 0.69 & 0.06 & 0.69 & 3.36  & 1.1E+02 &  0.93  &  0.38 \\
	\hline
	G355.2569+0.3646 &  69.1$\pm$0.2 &  6.7$\pm$0.5 & 2.85 & 0.06 & 2.85 & 13.39 & 4.0E+03 &  6.47  &  \\
	\hline
        G358.4116-0.3842 & -5.89$\pm$0.02  &  2.17$\pm$0.04 & 0.92 & 0.07 & 0.91 & 3.95  & 2.3E+02 &  2.62  &  \\
	\hline
	G358.5723-0.3058 & -3.94$\pm$0.04  &  3.4$\pm$0.1 & 1.46 & 0.06 & 1.46 & 6.66  & 4.4E+03 &  0.49  &  0.35 \\
	\hline
	G358.6771-0.4584 & -6.4$\pm$0.2  &  4.9$\pm$0.4 & 2.07 & 0.07 & 2.07  & 8.68  & 3.8E+03 &  6.03  &  \\
	\hline
	G359.0025+0.1684 & -2.7$\pm$0.1  &  1.3$\pm$0.2 & 0.54 & 0.06 & 0.54 & 2.50  & 6.8E+02 &  0.09  &  0.26 \\
	\hline
	G359.0305-0.2236 & -6.77$\pm$0.03  &  2.4$\pm$0.1 & 1.01 & 0.07 & 1.01 & 4.01  & 2.9E+03 &  0.34  &  0.17 \\
	\hline
	G359.2015+0.0068 & -1.0$\pm$0.1  &  1.6$\pm$0.3 & 0.67 & 0.07 & 0.67 & 2.80  & 1.7E+03 &  0.12  &  0.19 \\
	\hline
        G359.2513-0.0261 &  12.7$\pm$0.3 &  2.3$\pm$0.8 & 0.98 & 0.08 & 0.98 & 3.57  & 2.7E+03 &  0.70  &  0.08 \\
	\hline
	G359.3608-0.0077 & -104.9$\pm$0.7& 11.7$\pm$2.2 & 4.96 & 0.08 & 4.96 & 18.31 & 4.8E+04 &  13.29 &  \\
	\hline
	G359.4711-0.1175 & -1.7$\pm$0.1  &  1.9$\pm$0.3 & 0.79 & 0.07 & 0.79 & 3.15  & 1.6E+03 &  0.06  &  0.68 \\
	\hline	
	G359.4854-0.1237 & -2.0$\pm$0.1  &  1.3$\pm$0.3 & 0.55 & 0.07 & 0.55 & 2.23  & 5.9E+02 &  0.04  &  0.75 \\
	\hline
	G359.5086-0.0512 & -2.1$\pm$0.2  &  1.6$\pm$0.4 & 0.68 & 0.07 & 0.68 & 2.75  & 9.2E+02 &  0.07  &  0.58 \\
	\hline
	G359.5257-0.0096 & -1.4$\pm$0.1  &  2.0$\pm$0.2 & 0.86 & 0.07 & 0.86 & 3.39  & 9.9E+02 &  0.14  &  0.68 \\
	\hline
	G359.5372-0.0016 & -1.4$\pm$0.1  &  1.8$\pm$0.2 & 0.77 & 0.07 & 0.77 & 3.05  & 1.3E+03 &  0.12  &  0.38 \\
	\hline
	G359.5400-0.1638 &  0.8$\pm$0.6  &  2.5$\pm$1.4 & 1.05 & 0.07 & 1.05 & 4.33  & 1.6E+03 &  0.28  &  0.47 \\
	\hline
        G359.5454-0.0793 &  14.4$\pm$0.1 &  1.9$\pm$0.3 & 0.79 & 0.08 & 0.79 & 2.98  & 1.3E+03 &  0.16  &  0.34 \\
	\hline
	G359.5575-0.0933 &  14.3$\pm$0.3 &  2.3$\pm$0.6 & 0.97 & 0.08 & 0.97 & 3.63  & 3.9E+03 &  0.16  &  0.24 \\
	\hline
	G359.6100-0.1578 & -34.6$\pm$0.9 &  4.7$\pm$4.2 & 2.01 & 0.07 & 2.01 & 8.20  & 2.9E+03 &  4.92  &  \\
	\hline
	G359.6246+0.0235 & -76.0$\pm$1.0 & 11.2$\pm$1.1 & 4.74 & 0.07 & 4.74 & 18.57 & 2.0E+04 & 18.44  &  \\
	\hline
	G359.6454-0.0251 &  78.7$\pm$0.2 &  8.2$\pm$0.6 & 3.48 & 0.07 & 3.48 & 14.02 & 2.8E+04 &  6.41  &  \\
	\hline
	G359.6540-0.0064 &  -1.5$\pm$0.5 &  3.5$\pm$1.2 & 1.49 & 0.07 & 1.49 & 6.07  & 4.9E+03 & 0.57   &  0.30 \\
	\hline
	G359.6625-0.1137 & -33.5$\pm$0.4 &  3.9$\pm$1.6 & 1.65 & 0.07 & 1.65 & 6.46 & 1.9E+03 &  3.47  &  \\
	\hline
	G359.6749-0.1772 & -32.8$\pm$0.2 &  3.6$\pm$1.0 & 1.52 & 0.07 & 1.52 & 5.73 & 1.9E+03 &  3.97  &  \\
	\hline
	G359.6815-0.0842 &  14.7$\pm$0.1 & 13.6$\pm$0.3 & 5.76 & 0.07 & 5.76 & 23.51 & 1.1E+05 &  7.80  &  \\
	\hline
	G359.6930-0.0658 &  14.9$\pm$0.3 &  8.6$\pm$1.0 & 3.65 & 0.07 & 3.65 & 14.42 & 3.5E+04 &  6.38  &  \\
	\hline
        G359.6956-0.2986 &  9.62$\pm$0.02  &  1.5$\pm$0.1 & 0.64 & 0.08 & 0.64 & 2.25  & 1.1E+02 &  4.57  &  \\
	\hline
	G359.7119+0.0473 & -74.7$\pm$0.2 & 16.2$\pm$0.5 & 6.86 & 0.08 & 6.86 & 25.91 & 1.1E+05 & 13.87  &  \\
	\hline	
	G359.7338-0.1639 & -16.8$\pm$0.4 & 13.2$\pm$1.1 & 5.58 & 0.07 & 5.58 & 22.41 & 1.5E+04 & 35.06  &  \\
	\hline
	G359.7475-0.0359 &  14.5$\pm$0.2 & 14.2$\pm$0.4 & 6.04 & 0.08 & 6.04 & 22.18 & 7.3E+04 & 23.91  &  \\
	\hline
	G359.7513-0.1447 & -11.2$\pm$0.2 &  6.1$\pm$0.6 & 2.60 & 0.07 & 2.60 & 10.25  & 4.4E+03 &  4.69  &  \\
	\hline
	G359.7524+0.0426 & -67.0$\pm$0.5 &  9.26$\pm$1.0 & 3.93 & 0.08 & 3.93 & 14.81 & 1.8E+04 & 10.11  &  \\
	\hline
	G359.7781-0.0852 & -34.0$\pm$0.2 &  2.8$\pm$0.7 & 1.19 & 0.08 & 1.19 & 4.37 & 1.1E+03 &  2.63  &  \\
	\hline
        G359.8153+0.0262 & -62.9$\pm$0.4 &  7.3$\pm$1.7 & 3.11 & 0.08 & 3.11 & 11.58 & 4.3E+03 &  4.77  &  \\
	\hline
	G359.8170-0.0731 &  17.9$\pm$0.7 & 15.0$\pm$2.4 & 6.39 & 0.08 & 6.39 & 23.39 & 3.6E+04 &  8.99  &  \\
	\hline
        G359.8184-0.1286 &  5.9$\pm$0.4&  2.8$\pm$0.9 & 1.20 & 0.07 & 1.20 & 4.57  & 2.7E+03 &  0.27  &  0.47 \\
	\hline
	G359.8193+0.0365 & -61.7$\pm$1.0 &  8.9$\pm$4.8 & 3.77 & 0.08 & 3.77 & 14.22 & 1.7E+04 &  7.56  &  \\
	\hline
	G359.8514-0.3319 &  14.96$\pm$0.01 &  2.10$\pm$0.03 & 0.89 & 0.07 & 0.88 & 3.68  & 2.8E+02 &  3.07  &  \\
	\hline
	G359.8854-0.0763 &  15.9$\pm$0.1 &  21.7$\pm$0.2 & 9.23 & 0.07 & 9.23 & 39.43 & 4.2E+04 &  7.75  &  \\
	\hline
	G359.9207+0.0276 &  59.8$\pm$0.3 &  26.0$\pm$0.8 & 11.06 & 0.07 & 11.06 & 43.13 & 2.6E+05 & 44.81  &  \\
	\hline
\caption{Physical parameters of the 207 HMSC candidates. Source's names are listed in the first colunm. The 2nd to 10th column are the centroid velocities ($V_{\rm LSR}$), line widths ($\Delta V_{\rm FWHM}$), total velocity dispersion ($\sigma_{\rm v}$), thermal term velocity dispersion ($\sigma_{\rm th}$), non-thermal term velocity dispersion ($\sigma_{\rm nth}$), Mach number ($\mathcal{M_S}$), virial mass ($M_{\rm vir}$), virial parameters ($\alpha_{\rm vir}$), and required B-field to support against the collapse ($B$), respectively.}
\label{tab:parameter}
\end{longtable*}

\end{document}